\documentclass[aps,prb,twocolumn,superscriptaddress,citeautoscript]{revtex4}

\newcommand{\w}{\omega}

\renewcommand{\vec}[1]{{\bf #1}}

\usepackage{graphicx}
\usepackage{amsmath}

\graphicspath{{.}{./EPS/}}

\begin{document}

\title{Thermal Conductivity of Spin-$1/2$ Chains}

\author{E. Shimshoni}
\affiliation{Department of Mathematics--Physics, 
University of Haifa at Oranim, Tivon 36006, Israel}
\affiliation{Center for Materials Theory, Rutgers University,
Piscataway, NJ 08854--8019}

\author{N. Andrei} \affiliation{Center for Materials Theory,
Rutgers University, Piscataway, NJ 08854--8019} \author{ A. Rosch}
\affiliation{Institut f\"ur Theorie der Kondensierten Materie,
Universit\"at Karlsruhe, D-76128 Karlsruhe, Germany} 
\date{\today}

\begin{abstract}
We study the low-temperature transport properties of clean
    one-dimensional spin-$1/2$ chains coupled to phonons.  Due to the
    presence of approximate conservation laws, the heat current decays
    very slowly giving rise to an exponentially large heat
    conductivity, $\kappa \sim e^{T^\ast/T}$. As a result of an
    interplay of Umklapp scattering and spinon-phonon coupling, the
    characteristic energy scale $T^\ast$ turns out to be of order
    $\Theta_D/2$, where $\Theta_D$ is the Debye energy, rather than
    the magnetic exchange interaction $J$ -- in agreement with recent
    measurements in SrCuO compounds.  A large magnetic field $h$
    strongly affects the heat transport by two distinct
    mechanisms. First, it induces a {\it linear} spinon--phonon
    coupling, which alters the nature of the $T\rightarrow 0$ fixed
    point: the elementary excitations of the system are {\it composite
    spinon--phonon} objects.  Second, the change of the magnetization
    and the corresponding change of the wave vector of the spinons
    strongly affects the way in which various Umklapp processes can
    relax the heat current, leading to a characteristic fractal--like
    spiky behavior of $\kappa(T,h)$ as a function of $h$.
\end{abstract}


\maketitle

\section{Introduction and Main Results}
\label{sec:intro}

One--dimensional (1D) magnetic systems exhibit a variety of
interesting phenomena signifying their quantum many--body nature. They
have been, therefore, the subject of intense theoretical and
experimental study.  A great number of spin chain models has been
proposed and investigated, with various interaction ranges, spin
representations and anisotropies, as well as with coupling to other
degrees of freedom. The nearest neighbor spin - 1/2 Heisenberg model,
in particular, plays an important role being exactly
integrable\cite{bethe}, allowing a detailed analysis of its
thermodynamic properties \cite{takahashi}.  These properties are found
to be generic: they do not differ essentially from low energy
properties of other (non integrable) short-range spin chains.  On the
other hand, the Heisenberg model does not provide a generic
description of transport properties. Integrability entails the
existence of an infinite number of conservation laws, which in turn
imply a dissipationless transport\cite{zotos} of the elementary
excitations (spin-$1/2$ quantum solitons commonly named spinons) even
at {\em finite} temperatures, $T>0$ . As a result, measurable
dc-transport coefficients such as electric and thermal conductivities
are expected to be infinite.  In realistic systems described by spin
chains, e.g. materials consisting of weakly coupled chains (or
quasi--1D structures) of magnetically interacting ions, the
conservation laws of the ideal model are partially violated, and the
transport coefficients become finite. If the violation is soft one may
expect an unusually high thermal conductivity in such systems
attributed to spinon transport. Indeed, an experimental investigation
of heat transport in materials of the (Sr,Ca,La)CuO series have found
a considerable enhancement of the thermal conductivity in the
direction parallel to the chains \cite{spin-ex}. Other types of
systems involving quasi--1D, antiferromagnetic spin-$1/2$ chains such
as organic compounds \cite{org-sc} are in principle accessible to such
measurements, however at this point not much data is available
\cite{lorenz}.  Heat conductivity by magnetic excitations has also
been observed, for example, in the spin-Peierls compound\cite{ando}
CuGeO$_3$ or in ladder systems\cite{hess}.

In a recent experimental study of SrCuO compounds \cite{spin-ex-umk},
Sologubenko {\em et al.} report and analyze specific heat and thermal
conductivity data. The excess thermal conductivity ($\kappa_s$) along
the chains direction (obtained after subtracting the phonon
contribution) is identified as a contribution of the spinons, and its
dependence on the temperature $T$ is fitted to an empirical formula,
designed to account for scattering of the spinons by localized defects
as well as Umklapp processes. A phonon Umklapp mechanisms is suggested
as an interpretation of the exponential factor $\exp[T^\ast/T]$ which
describes $\kappa_s(T)$ in the range of temperatures $50$ K $\leq
T\leq 200$ K, with a characteristic temperature scale $T^\ast\sim 200$
K close to $\Theta_D/2$, where $\Theta_D$ is the Debye
temperature. However, this interpretation leaves open some questions:
As the system is at zero magnetic field (corresponding, in fermionic
language, to half filling) what inhibits the heat current relaxation
to proceed by low-energy spinon--spinon Umklapp scattering which is
not exponentially suppressed? Why is not the relevant energy scale to
the subleading relaxation processes the exchange interaction
$J/k_B\sim 2500 K$ rather than $\Theta_D/2 \sim 200$ K?  Another
interesting observation can be gleaned by conparing the plots of 
$ \ln \kappa(T)$ vs $1/T$ of the pure phonon contribution and of the
spinon contribution \cite{solo}.  We note that the slope of the former
is twice that of the latter. As pure phonon Umklapp processes are
goverened by $G$, the fundamental reciprocal lattice momentum, we need
to understand why it is $G/2$ that dominates the spinon processes.

We shall find that a rather subtle interplay of (approximate)
conservation laws and quantum dynamics underlies the 
experimentally observed heat conductivity, and a sophisticated
hydrodynamic field theoretic approach is necessary to fully account
for it. The arguments are rather general\cite{rosch}: When a system
possesses some conserved quantities $P$ these may ``protect'' the
current $J$ from degrading (this occurs when the cross-susceptibility
$\chi_{JP} \neq 0$) leading to a pure Drude peak and infinite d.c.
conductivity.  When the conservation of the pseudo-momenta $P$ is {\em
  softly} violated they will, instead, lead to very long time tails in
the decay of the current $J$, since states with a finite
pseudo-momentum $P$ typically carry also a finite current $J$ (when
$\chi_{JP} \neq 0$).  Hence the component of the current ``parallel''
to $P$, $J_{\parallel P} =(\chi^{\ }_{PJ} /\chi^{\ }_{PP}) P$ with
$\chi^{\ }_{JP}=\chi^{\ }_{J_{\parallel P}P}$, will therefore decay
exponentially slowly.  The presence of such approximately conserved
quantities leads then to a natural hydrodynamic description of the
system where a separation of fast and slowly decaying modes
takes place and a consistent scheme of calculation of the slow mode
conductivities can be carried out in terms of a matrix of decay rates of
these modes.

These features emerge when we study the transport properties of clean
1D spin chains in the framework of a generic model which accounts for
the coupling between spin excitations and lattice phonons. The model
is analyzed close to its fixed point Hamiltonian - the Luttinger
liquid - and the thermal conductivity $\kappa(T)$ is evaluated at low
but finite $T$. In this regime  the transport coefficients are
dominated by the slow relaxation of certain approximately conserved
currents due to irrelevant corrections to the fixed point Hamiltonian.
The most important correction terms of this sort, which are relatively
efficient in degrading the conserved currents of the integrable
Luttinger model, are found to be associated with Umklapp scattering
terms. These include pure spinon as well as spinon--phonon scattering
processes. In the absence of a magnetic field, the latter class is
shown to be dominant. Our detailed calculation of $\kappa(T)$
originating from Umklapp scattering of spinons by 3D phonons agrees
with the experimental results of Ref.~[\onlinecite{spin-ex-umk}].  In
particular, it explains the origin of the exponential factor
$\exp[T^\ast/T]$ with $T^\ast\sim \Theta_D/2$.

We then proceed to investigate the effects of a (large) external
  magnetic field $h$.  The induced magnetization and the corresponding change
  of the wave vector of the spinons strongly modifies the way how
  various Umklapp processes relax the heat current.  This leads to a
  fractal--like spiky behavior of $\kappa$ when plotted as a function
  of magnetization at fixed $T$, where the spikes occur at specific
  commensurate values of the magnetization.  Furthermore, the magnetic
  field induces a {\it linear} spinon--phonon coupling tunable by $h$.
  The coupling alters the nature of the fixed point: the elementary
  excitations of the system are {\it composite spinon--phonon}
  objects.  As a consequence of this mixing, the Umklapp
  processes are also modified and the relevant energy scale $T^*$ (again of
  the order of the minimum of $\Theta_D$ and $J$) depends smoothly on
  $h$. These effects are experimentally accessible in spin chains with
  relatively low magnetic exchange interaction $J$.

The paper is organized as follows: in Sec. \ref{sec:s-p-mod} we derive
the low energy model for the spinon system in the presence of coupling
to 3D phonons, and discuss the leading irrelevant corrections and
their significance for transport.  In Sec. \ref{sec:kappaTh0} we
present the calculation of the conductivity tensor by means of a
memory matrix approach, and derive expressions for the thermal
conductivity $\kappa$ as a function of the temperature $T$. Sec.
\ref{sec:kappaTh} is devoted to the study of $\kappa(T)$ in a finite
magnetic field $h$, where a linear coupling of spinons and 1D phonons
is accounted for within a Luttinger model. Our conclusions are
summarized in Sec. \ref{sec:conc}. 
 In Appendix~\ref{appendixAppr}
we emphasize that boundary conditions and finite size effects play an
important role in the presence of (approximate) conservation laws and
discuss what quantities are measured in a typical heat conduction
experiment.
Details of
the calculation of the memory matrix elements are given in
Appendix~\ref{appendixCalc}.  For convenience, throughout the paper we
adopt units where $\hbar=\mu_B=k_B=1$.

\section{Model for the weakly coupled spinon--phonon system}
\label{sec:s-p-mod}

 We wish to compute the low-temperature thermal conductivity of a
system consisting of a parallel array of long antiferromagnetic spin
chains embedded in a 3D lattice, and interacting with the lattice
phonons.  Let us begin by describing the spin system. A typical spin
chain model, with finite range interaction, is given by,
\begin{eqnarray}
H&=& \frac{1}{2}\sum_{i,j=1}^N  J_{ij}
\left(S_i^+S_{j}^- + S_i^-S_{j}^+\right)\nonumber \\
&+&\sum_{i,j=1}^N J_{ij}^z S_i^zS_{j}^z - h \sum_{i=1}^N S_i^z \; , \label{Hxxz}
\end{eqnarray}
where $S_i^\pm$, $S_i^z$ are spin-$1/2$ operators at lattice site $i$,
and $h$ is an external magnetic field applied along the
$z$-direction. The coupling is antiferromagnetic: $J_{ij} > 0$, translational
invariant: $J_{ij} = J_{i-j}$ and of finite range: $J_{i-j}=0$ for $
i-j > p$.

 The low energy dynamics of this class of models is described by a
 Luttinger liquid Hamiltonian. A fully fledged derivation would
 proceed via repeated RG transformations and yield, in principle, the
 fixed point Hamiltonian (the Luttinger liquid) as well as all
 irrelevant operators around it.  Instead of following this route, we
 shall employ a short cut and proceed via the Jordan--Wigner
 transformation allowing a fermionization of the spin degrees of
 freedom and subsequent bosonization. To illustrate it we consider the
 XXZ model corresponding to the choice of range $p=1$, but our
 conclusions are valid for a generic model. Begin by introducing a
 fermionic representation of the spin operators. In terms of spinless
 fermionic operators $\psi_i$ the spin operators can be expressed as
 \cite{affleck},
\begin{eqnarray}
S_i^-&=&\psi_i \exp\left[i\pi \sum_{j=1}^{i-1}n_j\right],\nonumber \\
S_i^z&=&:n_i:\; = \psi_i^\dagger\psi_i-\frac{1}{2}. \nonumber
\end{eqnarray}
The Hamiltonian $H_{xxz}$ is  mapped onto a model of interacting  
spinless fermions in $1D$, 
\begin{multline}
 H_{xxz} = -\frac{J}{2}\sum_i (\psi^\dagger_{i+1}\psi_i + \psi^\dagger_{i+1}\psi_i) \\
+ 
J_z \sum_i n_i n_{i+1} + h \sum_i n_i 
\end{multline}                            %
with $J_z$ determining the interaction strength
and $h$ playing the role of a chemical potential. The first term in $H_{xxz}$
corresponds to a kinetic energy term
\begin{equation}
H_k=-J\sum_k \cos (ka)\psi^\dagger_k\psi_k \nonumber
\end{equation}
(where $a$ is the lattice spacing), whose low energy excitations (the spinons)
are dominated by momenta $k$ in the close vicinity of 
the two Fermi points $\pm k_F$
with
\begin{equation}\label{kf-h}
k_F=\frac{\pi}{2 a} (1+M)
\end{equation}
where $M=2 \langle S_z \rangle \approx h/(\pi J)$ is the magnetization of the spin chain (normalized to $1$).


In the low energy limit the field operator  $\psi_j$    is 
 approximated as $\psi(x = ja ) \approx e^{i k_F x} 
\psi_R(x) +  e^{-i k_F x} \psi_L(x)$ where  the 
 right and left moving fields $\psi_R(x)$, $\psi_L(x)$ describe the low lying 
excitations, those  near 
$\pm k_F$ respectively. Using this expression in the Hamiltonian $H_{xxz}$
and keeping the leading terms only, one finds:
\begin{eqnarray}\label{HLLfer}
H_{xxz} \approx H_{LL} &=& -i (Ja)\int dx (\psi^\dagger_R\partial_x \psi_R -
\psi^\dagger _L\partial_x \psi_L) \nonumber \\
&&+ J_z \int dx (\rho_R^2+\rho_L^2+4\rho_R\rho_L)\; ,
\end{eqnarray}
where $\rho_{R/L}=\psi^\dagger_{R/L}\psi_{R/L}$.
The Luttinger Hamiltonian $H_{LL}$ thus obtained 
is conformally invariant. It is the
fixed point of $H_{xxz}$, but the coupling constant appearing in it are valid only to first order in $J_z/J$ and $ h/J$.

One may diagonalize $H_{LL}$ by changing to bosonic variables:
the field $\phi(x)$ and its conjugate $\Pi(x)$, 
satisfying $[\phi(x),\Pi(x^\prime)]=i\delta(x^\prime-x)$. The fermion fields
are then given by \cite{review}
\begin{eqnarray}
\psi_{R(L)}=\frac{1}{\sqrt{2\pi a}}
e^{i[\pm\phi-\theta ]} \; ,\nonumber
\end{eqnarray}
where $\theta$ is defined by $\partial_x\theta=\pi\Pi$. The 
advantage of the bosonic representation is that it allows to diagonalize the interaction
in $H_{xxz}$ by means of a Bogoliubov rotation leading to,
\begin{eqnarray}\label{HLL}
H_{LL}&=&v\int \frac{dx}{2 \pi}
\left( K (\pi\Pi)^2+\frac{1}{K} (\partial_x \phi)^2 \right) 
\end{eqnarray}
in which (to leading order in $|h|/ J$ and $|J_z|/ J$)
\begin{eqnarray}
v\approx\left(J+\frac{J_z}{\pi}\right)a\; , \quad K\approx\frac{1}{1+\frac{2J_z}{\pi J}}\; .
\end{eqnarray}
For arbitrary $|J_z|\leq J$ the Luttinger model Eq. (\ref{HLL}) still
captures the low energy physics of the spin chain, however its
derivation is more subtle. The exact Bethe Ansatz solution of the XXZ
model yields\cite{affleck} for $h=0$ (for $h\neq0$ see 
Ref.~[\onlinecite{cabra}])
\begin{eqnarray}\label{K-BA}
K=\frac{1}{2(1-\frac{1}{\pi}\cos^{-1}\left[\frac{J_z}{J}\right])}\; .
\end{eqnarray}
In particular, in the physically interesting case
of an isotropic Heisenberg antiferromagnet
 $J=J_z$  yielding $K=1/2$.
 
 As noted above the transport properties of the XXZ model are not
 generic. The infinite number of conserved charges which assure its
 integrability also lead to a pure Drude peak and infinite
 d.c.-conductivity even at finite temperature.  However the Luttinger
 liquid Eq. (\ref{HLL}) also describes (in the long wavelength limit)
 more complex spin chain structures as ladders, ``zigzag'' chains or
 in general chains with any finite range interaction (as long as no
 spin gap emerges) with the $J_{ij}$, $J_{ij}^z$ dependence of the
 parameters $v$ and $K$ given by model--specific combinations of the
 coupling coefficients.  The difference between integrable and
 non-integrable spin chains in the low energy limit is captured by the
 structure of the irrelevant operators around the Luttinger fixed
 point. In the latter case the irrelevant operators appear with
 generic coefficients.  Consideration of the irrelevant operators is
 of crucial importance for transport properties.  Non-integrable
 models are expected\cite{classical} to have a finite heat conductivity at $T>0$
 \cite{rosch,garst,meisner}. To compute it, however, the fixed point
 Hamiltonian $H_{LL}$ is insufficient by itself: it is
 translationally invariant and integrable, and therefore all the
 currents (e.g. spin current, heat current) described by it cannot
 degrade, leading to an infinite d.c. conductivity. One must add to
 the fixed point Hamiltonian all irrelevant operators around it, and
 compute the conductivity from the resulting effective low-energy
 Hamiltonian.  This implies, in passing, that the heat conductivity is
 a {\em singular} function of {\em irrelevant} perturbations,
 requiring us to recast perturbation theory in terms of a memory
 formalism (see below). We also note that a  number of recent
 studies\cite{infinite,gros} found an infinite heat conductivity in
 generic non-integrable models. We believe that these claims are a
 consequence of either numerical problems\cite{gros} or the neglect of
 certain classes of irrelevant perturbations\cite{infinite}.


We proceed to analyze the various irrelevant perturbations. In a
  generic model, all perturbations allowed by symmetry are generated
  when high energy modes are integrated out. The symmetries relevant
  for the following discussion are spin-rotation ${\mathcal R}_z$ around the
  z-axis, discreet translations by a lattice spacing ${\mathcal T}_a$, 
inversion ${\mathcal P}$ and
  time-reversal ${\mathcal T}$. In terms of the spinless fermions, ${\mathcal R}_z$
  guarantees charge conservation, ${\mathcal T}_a$ leads to momentum conservation
  up to reciprocal lattice vectors $G=2 \pi/a$ and the transformation
  rules under ${\mathcal P}$ and ${\mathcal T}$ are given by
\begin{eqnarray}
{\mathcal P}:&& \psi_L \to \psi_R, \psi_R \to \psi_L \\
{\mathcal T}: && \psi_L \to \psi_R^\dagger, \psi_R \to \psi_L^\dagger, i \to -i.
\end{eqnarray}

The operators consistent with the symmetries above will be further
divided into two classes, depending on their role in in transport
phenomena. The first class, $H^U$, consists of Umklapp operators
$H^U_{nm}$, describing processes where $n$ spinons are moved from the
right to the left Fermi point (and vice versa), possibly picking up
$m$ units of lattice momentum $G$. It is these processes that underlie
the degrading of the currents leading to finite conductivities. The
other class, $H_{irr}$, contains low-energy processes where the
number of spinons around each Fermi point remains conserved.  This
class includes corrections from band curvature, e.g. $\int
\psi^\dagger_R \partial^2 \psi_R$, or from finite-range
interactions. They do not affect the conductivities directly.  A
formal way to distinguish between the two classes is as
follows. Consider the two operators
\begin{eqnarray}
J_0 &=& N_R-N_L\; , 
\end{eqnarray}
where $N_R$ and $N_L$ are the total number of right and left moving spinons, respectively,
and the spinon translation operator 
\begin{eqnarray}\label{P-Ts}
P_{Ts}&=&\int dx [\psi_R^\dagger(- i \partial_x)
\psi_R+ \psi_L^\dagger(- i \partial_x) \psi_L]\; . 
\end{eqnarray}
These are among the infinite number of operators conserved by
$H_{LL}$, but play a special role in what follows.  As we show in the
next section, the conservation of certain linear combinations of $J_0$
and $P_{T}$ is minimally violated -- in comparison with all other
currents, the decay rates are exponentially small at low $T$. The
class of operators $H_{irr}$ consists of terms in the Hamiltonian
which conserve $N_R$, $N_L$ and are invariant under (continuous)
translations, hence commuting with both, 
$[H_{irr},J_0]=[H_{irr},P_{Ts}]=0$. The first
class, $H^U$, includes Umklapp operators which do not commute at least
with one of them.

\begin{widetext}
 For even $n$ the leading terms are of the form
\begin{eqnarray}\label{Unm-sc}
 H^U_{nm} &=& g^U_{nm} \int \!\! dx\, [ e^{i \Delta
k_{nm}x}\prod_{j=0}^{n} \psi_R^\dagger(x+j a) \psi_{L }(x+j a)
+h.c. ] = \frac{ g^U_{nm}}{(2 \pi a)^n} \int dx (
e^{i \Delta k_{nm} x}
e^{ i  2 n \phi(x) }+ h.c.)
\end{eqnarray}
where
\begin{equation}\label{knm}
\Delta k_{nm}= n 2 k_F-m G.
\end{equation}
Note that $\Delta k_{nm}$ depends on
the magnetization through $k_F$ and Eq. (\ref{kf-h}).
For odd $n$ and vanishing magnetic field $h$, 
time reversal invariance does not allow 
terms of the form (\ref{Unm-sc}), therefore leading perturbations are 
\begin{eqnarray}\label{Unm-scOdd}
 H^U_{nm} &=& g^U_{nm} \int dx [ e^{i \Delta
k_{nm}x}  (\psi_R^\dagger(x) \psi_{R }(x)+\psi_L^\dagger(x) \psi_{L}(x)) 
\prod_{j=1}^{n+1} \psi_R^\dagger(x+j a) \psi_{L }(x+j a)
+h.c.]  \\
&=& \frac{ g^U_{nm}}{(2 \pi a)^n} \int dx (
e^{i \Delta k_{nm} x}
e^{ i  2 n \phi(x) } \partial_x \phi(x)+ h.c.)\, .\nonumber 
\end{eqnarray}
\end{widetext}

Thus far we discussed in detail the low energy description of a single
spin chain. We now turn to consider the complete system (as investigated
experimentally e.g. by Sologubenko et al.\cite{spin-ex-umk})
  consisting of an array of spin chains
interacting with 3-dimensional acoustic phonons.

  The Hamiltonian of the system is
\begin{eqnarray}
H=H_{s}+H^{3D}_{p}  + H_{s,p}\; ,
\end{eqnarray}
where 
$H_s$ describes the spin array, $H^{3D}_{p}$ is the phonon Hamiltonian,
 and $H_{s,p}$ the interactions between
spins and phonons. The spin array Hamiltonian is simply given by a sum
of chains of the form we just discussed
\begin{eqnarray}
H_s = \sum_\alpha H^\alpha_{s} 
\end{eqnarray}
where $\alpha$ labels the spin chains (parallel to the $x$-axis) and
$H_s^\alpha=H_{LL}^\alpha+H_{irr}^\alpha+\sum_{nm} H^{U,\alpha}_{nm}$.

The Hamiltonian $H^{3D}_{p}$ describes the system of three dimensional
acoustic phonons to which the array of spin chains is coupled.  In the
following, we will consider mainly phonons describing deformations of
the lattice {\em parallel} to the chains which is chosen as the $x
-$direction.  The dynamics of these deformations, to be denoted by
$q$, is described by
\begin{eqnarray}\label{Hp3d}
H_{p}^{3D}=  \int \frac{d^3x}{2\pi} 
\left[ (\pi P)^2 + 
\sum_\mu  v_{\mu}^2 ( \partial_\mu q)^2 \right],
\end{eqnarray}
where $\mu$ denote the $x,y,z$ directions, $P$ and $q$ are
(appropriately normalized) canonical phonon momentum and coordinate
operators and $v_\mu$ are the sound velocities with $v_x=v_p$ and
$v_y=v_z=v_\perp$ assuming a tetragonal symmetry. Acoustic phonons
describing vibrations $\vec{q}_\perp$ perpendicular to the chains are 
omitted here but can easily be included. As we will
neglect in the following the weak phonon induced interactions between
different spin-chains, we will need only the propagator of phonons
along a single chain $\int d^2k_{\perp} G(\omega, \vec{k})=
\int d^2k_{\perp}\; 1/(\omega^2
+ \sum_\mu v_{\mu}^2 k_{\mu}^2) \sim \ln [ (\omega^2 + v_p^2
k^2)/\Theta_{D\perp}^2]$ where $\Theta_{D\perp}$ is the Debye
frequency perpendicular to the chains. 
The space-time form of the resulting  phonon propagator along a
chain at $T=0$ is $G_p(t,x) \sim 1/(x^2 + v_p^2 t^2)$.

The phononic and spin degrees of freedom couple in a variety of ways
which depend on the symmetries of the underlying lattice and on the
strength of spin-orbit coupling.  Assuming either inversion symmetry 
or weak spin-orbit coupling,
the dominant coupling arises from the dependence of the exchange
couplings on the distance of atoms $J_{i,j} = J(\vec{R_i-R_j})\approx J + a
(\partial_x q) J'+ O[\partial_x^2 q, (\partial_\mu q)^2, (\partial_\mu
\vec{q}_\perp)^2 ]$. 
  In the presence of a magnetic field, other couplings can become
  important as will be discussed in section~\ref{sec:kappaTh}.   In
 analogy to the classification of spinon-spinon interaction,
we classify the various spinon-phonon interactions
$H_{s,p}=H_{s,p}^{irr}+\sum_{nm} H^{U,s-p}_{nm}$ (again, generated by
integrating out high energy modes) into Umklapp and non-Umklapp operators 
by their commutation relations with the operators
$J_0$ and the translation operator $P_T$ of
{\em both} spinons and phonons
\begin{eqnarray}\label{PT-def}
J_0 &=& \sum_\alpha \int dx \,
[\psi_{R\alpha}^\dagger \psi_{R\alpha}- \psi_{L\alpha}^\dagger \psi_{L\alpha}]
\nonumber \\
P_T&=& - \int d^3 x P \partial_x q+\sum_\alpha P^\alpha_{Ts}
\label{PTall}
\end{eqnarray}
where $P^\alpha_{Ts}$ is the translation operator of spinons [Eq. (\ref{P-Ts})] on chain
$\alpha$. As we will show below, spinon-phonon couplings $H_{s,p}^{irr}$
which commute with both
$J_0$ and $P_T$, for example $\int (\partial\phi)^2\partial_x q$,
will not be able to relax the heat current completely. More important
are again Umklapp  terms  $H^{U,s-p}_{nm}$. For even $n$, leading contributions 
(in a given chain) are of the 
form 
\begin{equation}\label{Unm-sp}
H^{U,s-p}_{nm} = \frac{ g^{U,p}_{nm}}{(2 \pi a)^n} 
\int dx  [e^{i \Delta k_{nm} x}
e^{ i  2 n \phi } \partial_x q + h.c.]
\end{equation}
while for odd $n$ one obtains
\begin{equation}\label{Unm-spOdd}
 H^{U,s-p}_{nm} = \frac{ g^{U,p}_{nm}}{(2 \pi a)^n} 
\int dx  [e^{i \Delta k_{nm} x}
e^{ i  2 n \phi } (\partial_x q)( \partial_x \phi) + h.c.]
\end{equation}
where $\phi=\phi_\alpha$, the phonon field $q(\vec{R_\alpha},x)$ is
evaluated on the corresponding chain $\alpha$, and $\Delta k_{nm}$ is
given by (\ref{knm}).  Processes involving multiple phonons are
subleading.  
We would like to emphasize again, that {\em all}
the spinon-phonon couplings discussed above are irrelevant by
power--counting. But we have to keep them if they are the dominating
processes to relax the relevant approximate conservation laws.
Indeed, as will be shown in Sec. \ref{sec:kappaTh0}, the
spinon--spinon scattering terms Eq. (\ref{Unm-sc}), (\ref{Unm-scOdd})
for which $\Delta k_{nm}$ is finite are exponentially suppressed with
respect to the spinon--phonon terms in the physically relevant case,
where the Debye temperature $\Theta_D$ is much smaller than $J$.

\section{The Thermal conductivity for zero magnetic field}
\label{sec:kappaTh0}

We turn now to the computation of the transport properties of the spin
chains, assuming the parameters of the system to be compatible with the
SrCuO compounds studied in Ref. [\onlinecite{spin-ex-umk}]. Since the exchange
coupling $J$ in these materials is extremely high (over 2000 K),
magnetic field effects are negligible even at the strongest accessible
fields (of order a few tens of Tesla). Hence, throughout this section
we set $h=0$. We note that in other compounds where $J$ is much
smaller (e.g., organic spin chains), interesting magnetic field
effects should be observable, as will be shown in Sec.
\ref{sec:kappaTh}.  
We assume in addition the presence of inversion symmetry (or the absence of
 spin--orbit coupling), so that the spinons couple to
phonons only via $H_{s,p}=H_{s,p}^{irr}+\sum_{nm} H^{U,s-p}_{nm}$ [with
$H^{U,s-p}_{nm}$ given by Eqs. (\ref{Unm-sp}), (\ref{Unm-spOdd})].

The transport properties of the spin chains at low temperature - like
  the charge transport in (quasi-) one-dimensional
  metals\cite{rosch,garst,rosch02} - are governed by the approximate
  conservation of certain quantities $P_{nm}$ (to be called below
  ``pseudo-momenta'' ).  The exponentially slow decay of a given
  $P_{nm}$ will lead to an exponentially large heat conductivity for
  low $T$.  As already mentioned earlier, the reason is that states
  with a finite pseudo-momentum $P_{nm}$ typically carry also a finite
  heat current: the component of the heat current ``parallel'' to
  $P_{nm}$ will therefore decay exponentially slowly \cite{rosch}.

We now  proceed to identify the pseudo-momenta.
 Both $J_0$ and $P_T$ defined in (\ref{PTall}) decay
  rather slowly as they commute with all non-Umklapp terms
  $H_{LL},H_{irr},H_p^{3D},H^{irr}_{s,p}$. More important are
  certain linear combinations
\begin{equation}
P_{nm}=P_T+\frac{\Delta k_{nm}}{2 n} J_0
\end{equation}
which we call ``pseudo-momenta''  ($P_{n0}$ is the usual momentum
operator). The pseudo-momentum  $P_{nm}$ further commutes  with all Umklapp terms with the quantum
numbers $n$ and $m$ (and integer multiples $k n$, $k m$), namely:
\begin{eqnarray}\label{Tnm}
 [P_{nm}\!\!&,&\!\!H_p^{3D}+  H_{LL}+H_{irr}+H^{irr}_{s,p}]=0    
 \nonumber\\{}
 [ P_{nm}\!\!&,&\!\!H^U_{n,m}+H^{U,s-p}_{n,m}]=0=[ P_{nm} \;, \;H^U_{kn,km}+H^{U,s-p}_{kn,km}]\nonumber  
 \end{eqnarray}
For example, in the case of vanishing magnetic field 
considered in this section, 
where $k_F=G/4$ [see Eq. (\ref{kf-h})],
$P_{21}=P_T$ obviously commutes with all
translationally invariant terms with $\Delta k_{nm}=0$ and therefore
with all possible low energy processes.

 The fixed point Hamiltonian $H_{LL}$ being conformally invariant
possesses an infinite number of conservation laws. However, compared
to our pseudo-momenta these modes decay much faster when generic
perturbations are added since they do not commute with all the
(low-energy) terms which we have collected in $H_{irr}$ and
$H^{irr}_{s,p}$. They are therefore not important in the following
discussion.  The same argument applies to $H_{LL}+H^U_{21}$ which is
integrable or any other {\em particular} combination that happens to possess
conserved charges.

Having established the presence of the slowly decaying modes we now
turn to the question of how to calculate perturbatively transport
coefficients in a situation dominated by a few of these modes.  The
method of choice is the {\it memory matrix} approach
\cite{forster,woelfle,giamarchi,rosch} which is based on the idea that
while the conductivity is a highly singular function of the various
perturbations this is not the case for the matrix of decay-rates of
the slowest modes in the system.  The memory matrix approach is
formulated in a vector space of slowly decaying operators, spanned in
our case by $P_T$, $J_0$ and the heat current $J_Q$ as we want to
calculate the heat conductivity.  For convenience, we use instead of
$P_T$, $J_0$ and $J_Q$ the operators $J_T,J_s$ and $J_Q$ with:
\begin{equation}
J_T=v^2 P_T \quad \text{and}\quad J_s=vK J_0,
\end{equation}
where $J_s$ is the spin current. 
In  bosonized form
\begin{eqnarray}
J_T&=&-v^2\left[\int d^3x\, P \partial_x q+\sum_\alpha \int dx \,  
\Pi_\alpha \partial_x \phi_\alpha \right]\label{JT} \\
 J_s&=& vK\sum_\alpha\int dx \Pi_\alpha \; .\label{Js}
\end{eqnarray}

The heat current $J_Q=\int d^3x {\mathcal J}_{Qx}$ (along the chain
direction $x$) is determined from the continuity equation
$\partial_\mu{\mathcal J}_{Q\mu}+\partial_t{\mathcal H}=0 $
where ${\mathcal H}$ is the energy density. For low temperatures
it is sufficient to include only contributions from the fixed point, 
$\int d^3 x {\mathcal H} \approx H_{low}$ with 
\begin{equation}
H_{low}=H_p^{3D}+\sum_\alpha H_{LL}^\alpha \nonumber
\end{equation}
and one obtains
\begin{eqnarray}\label{JQ}
J_Q=-\int d^3x\, v_p^2 P \partial_x q
-\sum_\alpha \int dx \,  v^2\Pi_\alpha \partial_x \phi_\alpha \; .
\end{eqnarray}
Adding further contributions e.g. from Umklapp terms or band curvature does not affect
results to leading order.  
Note that the operators $J_Q$ and $J_T$ are intimately related -- in
fact, they differ by the relative weight of the spinon and phonon
degrees of freedom associated with the different velocities. 
However, there is a significant distinction between them: while $J_T\propto P_T$ 
remains conserved under all translational invariant corrections to $H_{low}$ 
[even those which mix spinons and phonons,  like
 $\int (\partial\phi)^2 \partial_x q$], $J_Q$ does not remain so.

We now set up the memory matrix formalism in the space spanned by the slow modes $J_T, J_s, J_Q$. To do so  we follow Ref.~[\onlinecite{forster}] 
and introduce a scalar product $(A|B)$ on the operators of the theory,
\begin{eqnarray}
\left(A(t)|B\right)&\equiv& T \int_0^{1/T} d\lambda
\left\langle A(t)^\dagger B(i \lambda) \right\rangle\; .
\end{eqnarray}
Then the dynamic correlation function of the operators $A$ and $B$ is
\begin{eqnarray}
 C_{AB} (\omega) &=& \int_0^{\infty} dt e^{i \omega t} \left(A(t)|B\right)\\
      &=& \left( A \left| \frac{i}{ \omega -{\mathcal L}} \right| B \right)\\
      &=& \frac{iT}{ \omega}
\int_0^{\infty} dt e^{i \omega t} \left\langle [A(t),B]\right\rangle-\frac{(A|B)}{i \omega}
\end{eqnarray}
with the Liouville operator ${\mathcal L}$ defined by ${\mathcal L}A = [H,A]$. 

In the space spanned by $J_T,J_s$ and $J_Q$ the matrix of 
conductivities is therefore given by
\begin{eqnarray}
\hat\sigma_{pq}(\w,T) =\frac{1}{T V} C_{J_p J_q} (\omega)\; ,
\end{eqnarray}
where $V$ is the volume of the system and $p,q$ are either of $T,s$ and $Q$. 
 The heat conductivity $\kappa$
is given by (c.f. Appendix~\ref{appendixAppr}) 
\begin{equation}\label{kappa-def}
\kappa(\w,T)=\frac{1}{T} \sigma_{QQ}(\w,T)
\end{equation}
and $\sigma_{ss}$ can be identified with the spin conductivity. The matrix of 
static susceptibilities can be written as
\begin{eqnarray}\label{chi-def}
\hat{\chi}_{pq}= \frac{1}{T V} (J_p|J_q)\; .
\end{eqnarray}

As argued above, the  matrix of conductivities $\hat{\sigma}$
has no good perturbative expansion. We
therefore  express it in terms of  a memory matrix $\hat{M}$ defined by
\begin{eqnarray} \label{sigma-t}
\hat\sigma(\w,T)&=&\hat{\chi}(T) \left(\hat{M}(\w,T)- i \w \hat{\chi}(T) \right)^{-1}
\hat{\chi}(T).  
\end{eqnarray}
and explicitly given as\cite{forster}
\begin{eqnarray}\label{M}
\hat{M}_{pq}(\w) =\frac{1}{T}
\left(\partial_t J_p \left| {\mathcal Q} 
\frac{i}{\w-{\mathcal Q}{\mathcal L}{\mathcal Q}} {\mathcal Q} 
\right| \partial_t J_q \right)\; .
\end{eqnarray}
Note that in the literature\cite{forster} the memory matrix is usually
defined as $\hat{M}\hat{\chi}^{-1}$. 
The operator  ${\mathcal Q}$ in Eq. (\ref{M}) is the projection operator 
on the space perpendicular to the slowly varying variables $J_p$,
\begin{eqnarray}\label{Q}
{\mathcal Q}=1-\sum_{pq} |J_p) \frac{1}{T} (\hat{\chi}^{-1})_{pq} (J_q|\; .
\end{eqnarray} 
This separation between fast and slow modes underlies the perturbative
expansion of $\hat{M}$ to which we now turn.

The perturbative evaluation of $\hat{M}$ is greatly simplified by the
observation that since $[H_{low},J_k]=0$ (for $k=s,T,Q$), the operators
$\partial_t J_k$ are already linear in perturbations around the low energy Hamiltonian 
$H_{low}$. Hence, when these coupling terms are included to leading order in 
perturbation theory, one can set ${\mathcal L}={\mathcal L}_{low}$ with 
${\mathcal L}_{low}=[H_{low},.]$ and ${\mathcal Q}=1$ in Eq. (\ref{M}). 
The expectation values in Eq. (\ref{chi-def}) are also computed with respect to
the low energy Hamiltonian $H_{low}$. Under these approximations, the
expression for the memory matrix Eq. (\ref{M}) can be written as
\begin{eqnarray}\label{MM}
\hat{M} = \frac{1}{T} \left[ \sum_{nm} (\hat{M}_{nm}+\hat{M}_{nm,s-p})\right] 
\end{eqnarray}
where $\hat{M}_{nm}$ and $\hat{M}_{nm,s-p}$ are matrices in the space of the slow modes with matrix elements given by,
\begin{eqnarray}\label{MMM} 
 M_{nm}^{pq}&\equiv& \frac{
\langle F^p;F^q \rangle^0_\w-
\langle F^p;F^q \rangle^0_{\w=0}}{i \w}\; ,  \\
 M_{nm,s-p}^{pq}&\equiv& \frac{
\langle F_{s-p}^p;F_{s-p}^q \rangle^0_\w-
\langle F_{s-p}^p;F_{s-p}^q \rangle^0_{\w=0}}{i \w}\; . \nonumber 
\end{eqnarray}
Here $F^p=i[J_p,H^U]$,  $\langle F^p;F^q\rangle^0_{\w}$ is the
retarded correlation function calculated with respect to $H_{low}$,
and similarly for $F_{s-p}^p=i[J_p,H^{U,s-p}]$ 
(the indices $n,m$ have been omitted for brevity). Note that all the matrices
$\hat{M}_{nm}$, $\hat{M}_{nm,s-p}$ are symmetric.
The static susceptibility matrix (for $aT\ll v_p$) is given by
\begin{eqnarray}\label{chi33}
\hat{\chi}&\approx& \left(
\begin{array}{ccc} 
2 vK/\pi & 0 & 0 \\
0 & \frac{\pi v T^2 }{3} & \frac{\pi v T^2 }{3} \\
0 & \frac{\pi v T^2 }{3} & \frac{\pi v T^2 }{3}
\end{array}
\right)
\end{eqnarray}
(where the matrix indices $p,q$ take the values  $s,T,Q$).

 We are mainly interested in the d.c.  thermal conductivity
 $\kappa(T)$ (Eq. (\ref{kappa-def}) at $\w=0$), which can be obtained
 from Eqs. (\ref{sigma-t}) and (\ref{MM}) through (\ref{chi33}) in the
 limit $\w\rightarrow 0$.  We find,
\begin{multline}\label{kappa-gen}
\kappa(T)  \approx  \frac{\pi^2 v^2T^3}{9}\left[(\hat{M}^{-1})_{TT}
+2(\hat{M}^{-1})_{QT}+(\hat{M}^{-1})_{QQ}\right]
\end{multline}
with 
\begin{eqnarray}\label{IinvM}
&&(\hat{M})_{pq}= \sum_{nml} (g^U_{nml})^2M^{pq}_{nl}(\Delta k_{nm},T)\; ,\\
&&M^{pq}_{nl}(\Delta k_{nm},T)\equiv \lim_{\w\rightarrow 0}M_{nml}^{pq}
\nonumber
\end{eqnarray}
For conciseness we introduced the index $l$: $l=0$ denoting
$M_{nm}^{pq}$, and $l=1$ denoting $M_{nm,s-p}^{pq}$ (and similarly for
the coupling constants $g^U$). Note that we have only retained
contributions of Umklapp operators in $M^{QQ}$.  There are further
contributions arising from $H^{irr}_{s,p}$ which turn out to be
subleading for vanishing magnetic field and are therefore omitted
here.  They are, however, important in the case of a finite
magnetization, cf.  section~\ref{sec:kappaTh}.
  
The characteristic Luttinger liquid behavior of the spinon system is
reflected by the functional dependence of $M_{nl}^{pq}(\Delta k,T)$ on
$T$ and $\Delta k$.  Approximate expressions for these functions can
be obtained analytically in the high $T$ limit, $T \gg v |\Delta k|$, or
the low $T$ limit, $T \ll v_{p} |\Delta k|$, (see Appendix~\ref{appendixCalc}
for a
detailed calculation).  We first note that the conservation law
(\ref{Tnm}) implies a trivial relationship between $F^s$ and $F^T$,
and consequently for any $T$ and $\Delta k$
\begin{eqnarray}\label{MsT}
M_{nl}^{sT}(\Delta k,T)&=& - \frac{v \Delta k}{2nK} M_{nl}^{ss}(\Delta k,T) 
\; ,\nonumber \\
M_{nl}^{TT}(\Delta k,T)&=& \frac{v^2(\Delta k)^2}{4n^2K^2} M_{nl}^{ss}(\Delta k,T)\; , \\
M_{nl}^{QT}(\Delta k,T)&=& - \frac{v \Delta k}{2nK}M_{nl}^{Qs}(\Delta k,T)\; .
\nonumber
\end{eqnarray}
We therefore need to compute directly only three types of functions: 
$M_{nl}^{ss}$, $M_{nl}^{Qs}$ and $M_{nl}^{QQ}$. In the high $T$ limit we get
\begin{eqnarray}\label{MhighT}
M_{nl}^{ss}(\Delta k,T)&\sim & T^{2(n^2K+l)-3} \nonumber \\
M_{nl}^{Qs}(\Delta k,T)&\sim &M_{nl}^{ss}(\Delta k,T) \Delta k \\ 
M_{nl}^{QQ}(\Delta k,T)&\sim &M_{nl}^{ss}(\Delta k,T)T^2 \nonumber \; .
\end{eqnarray} 
More interesting is the low $T$ limit $T\ll v_p |\Delta k|$, 
in which we find
\begin{eqnarray}\label{MlowT}
M_{nl}^{ss}(\Delta k,T)&=& (2nKv)^2 M_{nl}(\Delta k,T)\; , \nonumber \\
M_{nl}^{Qs}(\Delta k,T)&\sim& \Delta k M_{nl}^{ss}(\Delta k,T)\; , \\
M_{nl}^{QQ}(\Delta k,T)&\sim& (\Delta k)^2 M_{nl}^{ss}(\Delta k,T)
\nonumber
\end{eqnarray}
where in the last line we have used the additional assumption
$(T/v |\Delta k|) \ll (v_p/v) $. The expressions for
$M_{nl}(\Delta k,T)$ are the following: 
For even $n$
\begin{eqnarray}\label{MnlowT0}
M_{n0}(\Delta k,T)\approx e^{ -\frac{v |\Delta k|}{2 T}}
\frac{a^{2-2n}}{ \pi^2 \Gamma^2(n^2K/2) v T} 
  \left( \frac{ a \Delta k}{2} \right)^{n^2K-2}
\end{eqnarray}
and 
 
\begin{widetext}
\begin{eqnarray}\label{MnlowT}
M_{n1}(\Delta k,T)&\approx & A
\frac{a^2v}{(2\pi a)^{2n}v_p^2T(\Delta k)^2} 
\left(\frac{v_p}{v}\right)^{2(n^2K-1)} 
\left(\frac{aT}{2 v}\right)^{2n^2K}
\left(\frac{a|\Delta k|}{2}\right)^{2}
\exp \left[-\frac{v_p |\Delta k|}{ 2T} \right]  
\end{eqnarray}
($A$ is a numerical factor). For odd $n$, $M_{nl}$  are given by the above expressions
multiplied by a factor $\sim(\Delta k)^2$ (for $l=0$) or $\sim(T/v)^2$ (for $l=1$),
in particular
\begin{eqnarray}\label{MnlowTOdd}
M_{n1}(\Delta k,T)&\approx & \tilde{A}
\frac{a^2T}{(2\pi a)^{2n}v_p^2v(\Delta k)^2} 
\left(\frac{v_p}{v}\right)^{2(n^2K-1)} 
\left(\frac{aT}{2 v}\right)^{2n^2K}
\left(\frac{a|\Delta k|}{2}\right)^{2}
\exp \left[-\frac{v_p |\Delta k|}{ 2T} \right].  \nonumber 
\end{eqnarray}
\end{widetext}
Note that the above exponential factors are always dictated by the
{\it smallest} of the velocities involved, and in our case
$v_p=v_{min}=min\{v,v_p\}$. The physical origin of this behavior is
that the minimal energy cost of a process involving a momentum
transfer of $\Delta k$ is associated with initial and final states of
the elementary excitations with energy $v_{min}\Delta k/2$ each. Since
in the system of interest to us $v_p\ll v$, the exponential factor in
(\ref{MnlowT0}) dramatically suppresses the pure spinon contribution
to the sum in Eq. (\ref{IinvM}) (in particular, for $|\Delta k|\sim
1/a$ the exponent becomes $\sim -(J/T)$).  However, among the
particular Umklapp scattering terms for which $\Delta k=0$ (and hence
the `high $T$' limit (\ref{MhighT}) applies), the spinon--spinon
process dominates as it contributes the leading power of $T$.

We now focus our attention on the low $T$ behavior of the thermal
conductivity Eq. (\ref{kappa-gen}).  Eq. (\ref{kf-h}) implies that for
$h=0$, $k_F=\pi/2a=G/4$.  This is a particular, commensurate value of
the filling $2k_F/G$, in which case the Umklapp term $n=2$, $m=1$ does
not involve a momentum transfer, i.e. $\Delta k_{21}=0$. Due to the
exponential factor in Eqs. (\ref{MnlowT0}) and (\ref{MnlowT}), the sum
in Eq. (\ref{IinvM}) is strongly dominated by terms with a minimal
$\Delta k$. In particular, the leading contribution to $M^{ss}$ and
$M^{QQ}$ is the single term $n=2$, $m=1$, $l=0$ corresponding to
$\Delta k_{21}=0$, where $M_{20}^{ss}(0,T)$ and $M_{20}^{QQ}(0,T)$ are
given by Eq. (\ref{MhighT}).  The other matrix elements vanish for
$\Delta k=0$. In fact, the vanishing of $M_{21}^{pT}$,
$M_{21}^{Tq}$ (for any $p,q=s,T,Q$) reflects the fact that $J_T\propto
P_{21}$ commutes with all low energy terms in the Hamilitonian. Their
leading contribution is therefore associated with the next smallest
$\Delta k$, i.e.  the term $n=1$, $m=0$ and $l=1$. These are given by
the low $T$ approximation (\ref{MnlowTOdd}) with $n=1$, $\Delta
k=G/2$. As a result
\begin{eqnarray}
(\hat{M}^{-1})_{TT}\approx \frac{1}{M_{11}^{TT}(G/2,T)}
\end{eqnarray}
which is {\it exponentially diverging}. In contrast,
$(\hat{M}^{-1})_{QQ}$ and $(\hat{M}^{-1})_{QT}$ are inversely
proportional to $M_{QQ}$, associated with the relatively fast
short-time relaxation rate of the heat current $J_Q$. As a consequence
they depend {\it algebraically} on $T$ and hence are exponentially
suppressed compared to $\hat{M}^{-1}_{TT}$. Inserting into
Eq. (\ref{kappa-gen}) this yields
\begin{eqnarray}\label{kappa-h0}
\kappa(h=0) &\approx & \kappa_0 \left(\frac{T}{T^\ast}\right)^{2(1-K)}
\exp  \left[\frac{T^\ast}{ T}\right]\; ,
\end{eqnarray}
with
\begin{eqnarray}
T^\ast & = & \frac{v_p G}{4}\, 
\end{eqnarray}
and $\kappa_0$ depending on the parameters of the spinon--phonon
system and the typical Umklapp scattering strength $g^2$
($\kappa_0\sim g^{-2}$). 

How does this compare to experiments? An exponential behavior of the
spin contribution to the heat conductivity has indeed been observed by
Sologubenko {\it et al.}\cite{spin-ex-umk} in SrCu$_2$ and
Sr$_2$CuO$_3$ for temperatures above $50\,$K (below which scattering
from defects seems to become important). It was emphasized by the
authors that $T^*$ is of the order $\Theta_D/2$, where $\Theta_D$ is
the Debye temperature.  A precise comparison to our result would
  require a detailed knowledge of the phonon velocities in these
  systems. However, if we neglect for simplicity all anisotropies of
  the phonons, the Debye temperature is given by $\Theta_D \approx
  v_p(6\pi^2/a^3)^{1/3}\approx 0.6 \, v_p G$ and therefore $T^\ast
  \approx 0.4\, \Theta_D$ in very good agreement with the experimental
  observation.  
  
  In (\ref{kappa-def}), we have defined $\kappa$ to be determined from
  an energy-current correlation function. However, in the presence of
  exact or approximate conservation laws and for finite systems it is
  far from obvious that this is the quantity measured in a typical
  heat transport experiment. For example, if spin is exactly
  conserved, boundary conditions will imply that no spin-current will
  flow through the surface of the sample and therefore the heat
  current in the experiment has to be calculated under the boundary
  condition of vanishing spin-current. As explained in detail in
  Appendix~\ref{appendixAppr}, this implies that $\sigma_{QQ}$ in
  Eq.~(\ref{kappa-def}) has to be replaced by
  $\sigma_{QQ}-\sigma_{Qs}^2/\sigma_{ss}$. However, we consider in this
  paper a different limit, assuming that the sample is much longer
  than typical length scales on which e.g. the spin does decay. Under
  these assumptions, Eq.~(\ref{kappa-def}) is indeed valid -- see
  Appendix~\ref{appendixAppr} for details.

{\em A note added on July 12, 2005: an error in the form of the phonon 
propagator lead to a wrong power-law prefactor in Eq. (47) -- the power 
$2(1-K)$ should be replaced by $-2K$. See Appendix C for details.}

\section{Effects of a finite magnetic field}
\label{sec:kappaTh}

 We now consider the effect of a finite magnetic field $h$ on the heat
conductivity of the spin chain. The field will have two main effects:
the first effect will be to modify $k_F$ [see Eq. (\ref{kf-h})] and
hence $\Delta k_{nm}$ leading to a fractal--like structure of the
conductivity as a function of the magnetic field: as $h$ is varied the
system passes from incommensurable to commensurable values (for which
$\Delta k_{nm} =0$ for certain values of $n,m$) leading to a strong variation of the conductivity
(see below). Clearly for this effect to be measurable, the spin--spin
coupling $J$ cannot be too large: to be observable with accessible
fields, $J$ needs to be of the order of a few tens degrees Kelvin.
The second effect of a finite $h$ is to induce linear phonon--spin coupling
by the field, which alters the fixed point Hamilitonian of the
system. Such a coupling is possible as the magnetic field breaks time
reversal invariance ${\mathcal T}$. (A similar coupling can arise as a
consequence of spin--orbit interaction in crystals without inversion
symmetry, even when $h=0$).  For finite $h$ the linear coupling
 arises from terms of the form
$(\partial_x q)\vec{S}_i \vec{S}_j \approx (\partial_x q) M \delta
S_z$ where $M=2 \langle S_z \rangle$ is the magnetization and $\delta
S_z=S_z- \langle S_z \rangle$.

  To analyze this case, we focus for simplicity on a strictly 1D
geometry considering only longitudinal phonons traveling along the
chain direction. We note, however, that much of our forthcoming
predictions are expected to be qualitatively applicable to spin chains embedded in
higher dimensional systems as well.

For 1D phonons the free Hamiltonian Eq. (\ref{Hp3d}) reduces to
\begin{eqnarray}\label{Hp}
H_{p}=v_p\int \frac{dx}{2\pi} \left[ (\pi P)^2 + (\partial_x q)^2 \right].
\end{eqnarray} 
The normalization of $P$ and $q$ here is chosen differently than in Eq. (\ref{Hp3d}),
so that their dimensions are the same as $\Pi$ and $\phi$, respectively.
It thus has the form of a Luttinger liquid with velocity $v_p$ and 
Luttinger parameter $K=1$.
To this we add a spinon--phonon coupling term of the form
\begin{eqnarray}\label{Hs-p}
H_{s-p}= -u_0\int \frac{dx}{\pi}\partial_x \phi\partial_x q \; .
\end{eqnarray}
At finite magnetic field, $u_0$ grows linearly with the magnetization
and as a consequence can be controlled. For $h=0$ and in the absence
of inversion symmetry $\cal P$ similar terms which couple linearly to
$P$ rather than $\partial_x q$ arise from spin orbit coupling. 
They also give rise to a mixing of modes (the roles of $P$
and $\partial_x q$ can be interchanged in the analysis below).

A term of the form (\ref{Hs-p}) leads to new eigenmodes of mixed
 spinon--phonon excitations.  The Hamiltonian
\begin{equation}\label{H_0}
H^*=H_{LL}+H_p+H_{s-p}
\end{equation}
is still scale invariant and is the fixed point of the coupled
spinon-phonon system.
We turn to  diagonalize it. It  is useful to define the free boson fields
$\tilde\phi=\phi/\sqrt{K}$ and $\tilde\Pi=\sqrt{K}\Pi$, in terms
of which Eq. (\ref{H_0}) can be written as
\begin{eqnarray}\label{H_0-new}
H^*&=&\int \frac{dx}{2\pi}v[(\pi\tilde\Pi)^2+
(\partial_x \tilde\phi)^2] \\ 
\,&+&\int \frac{dx}{2\pi}\left\{v_p[(\pi  P)^2+(\partial_x  q)^2]
-2u\partial_x\tilde\phi\partial_x  q \right\}\nonumber  
\end{eqnarray}
($u=u_0\sqrt{K}$). We then diagonalize $H^*$ using the
 transformation \cite{orignac}, 
\begin{eqnarray}\label{phi1phi2}
\left(\begin{array}{cc} \tilde\phi \\  q \end{array}\right)
= \left(\begin{array}{cc} C & -S(v/v_p)^{1/2} \\ 
  S(v/v_p)^{1/2} & C \end{array}\right)
\left(\begin{array}{cc} \phi_1 \\ \phi_2 \end{array}\right) 
\end{eqnarray}
and similarly for the canonical momenta
\begin{eqnarray}\label{pi1pi2}
\left(\begin{array}{cc} \tilde\Pi \\  P \end{array}\right)
= \left(\begin{array}{cc} C & -S(v_p/v)^{1/2} \\ 
  S(v_p/v)^{1/2} & C \end{array}\right)
\left(\begin{array}{cc} \Pi_1 \\ \Pi_2 \end{array}\right) \; ,
\end{eqnarray}
where (assuming $v>v_p$)
\begin{eqnarray}\label{defCS}
C\equiv \frac{1}{\sqrt{2}}\left[1+\frac{v^2-v_p^2}{U^2}\right]^{1/2}\nonumber \\
S\equiv \frac{1}{\sqrt{2}}\left[1-\frac{v^2-v_p^2}{U^2}\right]^{1/2} \\
U^2\equiv \left[(v^2-v_p^2)^2+u^2vv_p\right]^{1/2} \; .\nonumber
\end{eqnarray}
The transformation (\ref{phi1phi2}), (\ref{pi1pi2}) is symplectic in
order to preserves the canonical commutators,
$[\phi_\nu(x),\Pi_{\nu^\prime}(x^\prime)]=i\delta_{\nu
\nu^\prime}\delta(x-x^\prime)$.  The resulting Hamiltonian $H^*$ takes
the diagonal form
\begin{eqnarray}\label{H0final}
H^*&=&\int \frac{dx}{2 \pi}
\sum_{\nu=1,2}  v_\nu\left( K_\nu (\pi
\Pi_\nu)^2+\frac{1}{K_\nu} (\partial_x \phi_\nu)^2 \right)
\end{eqnarray}
where
\begin{multline}
\frac{v_1}{K_1}=\frac{1}{2v}\left[v^2+v_p^2+U^2\right]\; ,\quad  
\frac{v_2}{K_2}=\frac{1}{2v_p}\left[v^2+v_p^2-U^2\right]\\
 v_1K_1 = v,\qquad v_2K_2 = v_p. 
\end{multline}
The strongest mixing (for a given $u$) occurs when $v\sim v_p$, which is realized
in spin chain systems provided the exchange
interaction $J$ is not too large, and comparable to $\Theta_D$. We then define
${\bar v}=(v+v_p)/2$, $\delta v=v-v_p$, where $\delta v\ll {\bar v}$.
Assuming in addition $u\ll {\bar v}$, we obtain 
\begin{eqnarray}
v_1&\approx &{\bar v}[1+\Delta(h)] \, ,\;\; \;\;\; \quad  v_2\approx {\bar v}[1-\Delta(h)]\, ,  \\
K_1&\approx &1+ \frac{\delta v}{2{\bar v}}-\Delta(h)\, ,\;\; 
K_2\approx 1- \frac{\delta v}{2{\bar v}}+\Delta(h)\, ,\nonumber
\end{eqnarray}
where
\begin{eqnarray}\label{Delh}
\Delta(h)=\frac{[4(\delta v)^2+u^2]^{1/2}}{4{\bar v}}
\end{eqnarray}
which depends on the magnetic field $h$ via the coupling $u$.

We proceed to evaluate the heat transport coefficient following the
memory matrix method described in section~\ref{sec:kappaTh0}. The
linear coupling modifies the low-energy heat current $J_Q$, which is
now defined with respect to the elementary degrees of freedom of $H^*$
rather than $H_{LL}+H_p$. In terms of the two eigenmodes of $H^*$,
$J_Q$ is given by
\begin{eqnarray}\label{JQ12}
J_Q=-\int dx\sum_{\nu=1,2} v_\nu^2 \Pi_\nu\partial_x \phi_\nu\; .
\end{eqnarray} 
In analogy with Eq. (\ref{JT}), $J_T$ (now defined as $J_T=\bar{v}^2P_T$)
is given by
\begin{eqnarray}\label{JT12}
J_T=-\bar{v}^2\int dx\sum_{\nu=1,2}  \Pi_\nu\partial_x \phi_\nu\; .
\end{eqnarray}  
The derivations of the memory matrix and the static susceptibility
proceed in the same way as in Sec. \ref{sec:kappaTh0}.  In particular,
the memory matrix $\hat{M}$ is generally given by Eq. (\ref{MM}) and
by (\ref{IinvM}) in the $\w\rightarrow 0$ limit. The static
susceptibility matrix $\hat{\chi}$ is given by an expression nearly
identical to (\ref{chi33}), except that in $\hat{\chi}_{TT}$,
$\hat{\chi}_{TQ}$ and $\hat{\chi}_{QQ}$ the velocity $v$ is replaced
by $2\bar{v}$ (to leading order in $\Delta(h)$). The most prominent
modification compared to the $h=0$ case is indicated in the
spinon--spinon contribution to $\hat{M}$ (i.e., the $l=0$ terms): as a
result of the composite nature of the elementary excitations, it is no
longer a pure spin contribution, and is dominated at low $T$ by the
same exponential factor as the $l=1$ terms.  Since the latter involve
a higher power of $T$, $M_n^{pq}(\Delta k_{nm},T)\equiv M_{nm0}^{pq}$
actually dominates for any $\Delta k_{nm}$. We therefore neglect the
$l=1$ terms, and Eq. (\ref{IinvM}) becomes
\begin{eqnarray}\label{Msum}
\hat{M}(T)= \sum_{nm} (g^U_{nm})^2 \hat{M}_{n}(\Delta k_{nm},T)\; .
\end{eqnarray}

Similarly to Eq. (\ref{MsT}), we find
\begin{eqnarray}\label{MsT12}
M_{nl}^{sT}(\Delta k,T)&=& - \frac{\bar{v}^2 \Delta k}{2nKv} M_{nl}^{ss}(\Delta k,T) 
\; ,\nonumber \\
M_{nl}^{TT}(\Delta k,T)&=& 
\frac{\bar{v}^4(\Delta k)^2}{(2nKv)^2} M_{nl}^{ss}(\Delta k,T)\; , \\
M_{nl}^{QT}(\Delta k,T)&=& - \frac{\bar{v}^2 \Delta k}{2nKv} M_{nl}^{Qs}(\Delta k,T)\; .
\nonumber
\end{eqnarray}
The calculation of the functions $M_n^{pq}(\Delta k,T)$ (for $p,q$
denoting either of $s,Q$) is essentially the same as in the $h=0$ case
(see Appendix~\ref{appendixCalc}).  In the high $T$ limit ($T \gg v_1
|\Delta k|$) we get
\begin{eqnarray}\label{MhighT12}
M_n^{ss}(\Delta k,T)&\sim & T^{2(\alpha_n+\beta_n)-3} \nonumber \\
M_n^{Qs}(\Delta k,T)&\sim &M_n^{ss}(\Delta k,T) \Delta k \\ 
M_n^{QQ}(\Delta k,T)&\sim &M_n^{ss}(\Delta k,T)T^2 \nonumber \; ,
\end{eqnarray} 
where
\begin{eqnarray}\label{alpbet}
\alpha_n &\equiv & n^2KK_1C^2\; ,\; \; \beta_n\equiv n^2KK_2S^2(v/v_p)
\end{eqnarray}
and $C$, $S$ are defined in Eq. (\ref{defCS}). The low $T$ limit corresponding to
$T\ll v_2 |\Delta k|$ and $T/(v_1 |\Delta k|) \ll  S\ll 1$ yields
\begin{eqnarray}\label{MlowT12}
M_n^{ss}(\Delta k,T)&=& (2nKv)^2 M_n(\Delta k,T)\; , \nonumber \\
M_n^{Qs}(\Delta k,T)&\sim& \Delta k M_n(\Delta k,T) \; , \\
M_n^{QQ}(\Delta k,T)&\sim& (\Delta k)^2 M_n(\Delta k,T) \nonumber
\end{eqnarray}
with
\begin{eqnarray}\label{MnlowT12}
&& M_n(\Delta k,T)\approx  \frac{Aa^2}{(2\pi a)^{2n}{\bar v}T(\Delta k)^2
[\Delta(h)]^{\alpha_n}(2\beta_n)^{2\beta_n}} \nonumber  \\
&& \times   \left(\frac{aT}{2 \bar{v}}\right)^{2\alpha_n}
\left(\frac{a|\Delta k|}{2}\right)^{2\beta_n}
\exp \left[-\frac{v_{2} |\Delta k|}{ 2T} \right]\, .  
\end{eqnarray}
Here $A$ is a numerical factor, $\Delta(h)$ is defined in Eq. (\ref{Delh}) 
and we have used the fact that $v_2=\min\{v_1,v_2\}$. 
Note that since time--reversal symmetry is broken at finite $h$, Eq. (\ref{Unm-sc})
is the  leading Umklapp term for arbitrary $n$, and hence Eq. (\ref{MnlowT12}) hold for
both odd and even $n$.

Inserting Eqs. (\ref{MsT12}) through (\ref{MnlowT12}) into
(\ref{Msum}) we get the dominant contributions to  most elements of the matrix
$\hat{M}(T)$. An exception is $\hat{M}_{QQ}$, which includes
additional corrections neglected in the above approximations: these
are associated with irrelevant perturbations such as $\int
(\partial\phi)^2\partial_x q$, which do not commute with the heat current
$J_Q$.  These lead to contributions to $\hat{M}_{QQ}$ which are power
law in $T$. 
In the case of a finite magnetization, when all the Umklapp terms are 
either suppressed exponentially [Eq. (\ref{MnlowT12})]
or by large powers of $T$ (\ref{MhighT}),
these corrections cannot be neglected. 
Hence, at low $T$ one always gets $\hat{M}_{QQ}\gg
\hat{M}_{TT},\hat{M}_{TQ}$ and $\kappa$ is dominated
 by $\hat{M}^{-1}_{TT}$. Using that
$\hat{M}_{QQ}$ is exponentially larger than the smallest
eigenvalue of the matrix $\hat{M}$, we obtain 
with exponential precision $\hat{M}^{-1}_{TT}\approx 
\hat{M}_{ss}/(\hat{M}_{ss} \hat{M}_{TT}-\hat{M}_{sT}^2)$ and therefore
for the thermal conductivity $\kappa$
\begin{eqnarray}\label{kappa-lowTh}
\kappa(T) &\approx &\frac{4\pi^2{\bar v}^2 T^3}{9}\hat{M}^{-1}_{TT}(T)
\approx \frac{4\pi^2 T^3}{9{\bar v}^2}
\frac{\sum_{nm}M_{nm} n^2}{{\mathcal D}} \; ,
\end{eqnarray}
with 
\begin{eqnarray}
{\mathcal D}\approx \frac{1}{2}\sum_{nm}\sum_{n^\prime m^\prime}
M_{nm}M_{n^\prime m^\prime }(n\Delta k_{n^\prime m^\prime}
-n^\prime\Delta k_{nm})^2\;  \label{Den-lowT}
\end{eqnarray}
and $M_{nm}$ as an abbreviation for $(g^U_{nm})^2M_{n}(\Delta k_{nm},T)$.

How does the conductivity $\kappa$ depend on $T$ and $h$? As discussed
  above, a magnetic field $h$ leads to a linear spinon-phonon coupling
  and therefore tunes the parameters $\Delta(h)$ and $v_{2}$ in Eq.
  (\ref{MnlowT12}). However, for large fields, of the order $h\sim J$, another
  effect is even more important: the finite magnetization $M$ induces
  a shift of the Fermi momentum $k_F$ according to Eq.
  (\ref{kf-h}). The filling $2k_F/G=\frac{1+M}{2}$ is set to an
  arbitrary, generally irrational value, and can be tuned continuously
  by varying $h$. Upon tuning $k_F$, the characteristic momentum
  transfer $\Delta k_{nm}$ (\ref{knm}) associated with an Umklapp
  process $H^{U}_{nm}$ is modified accordingly.  As Umklapp processes
  at low $T$ are suppressed exponentially by $e^{-v_2 \Delta k_{nm}/(2
  T)}$, a change in $\Delta k_{nm}$ modifies exponentially the
  contribution of $H^{U}_{nm}$ to the various relaxation rates.
  
  We first analyze $\kappa$ for low $T$ and an almost commensurate
  magnetization with $M\approx 2 \frac{m_0}{n_0}-1$ or $k_F\approx
  \frac{G}{2} \frac{m_0}{n_0}$ where $m_0$ and $n_0$ are (small)
  integer numbers. In this situation, $H^U_{n_0 m_0}$ is the strongest
  Umklapp term as, according to (\ref{knm}), $\Delta k_{n_0 m_0}
  \approx 0$. However, due to the approximately conservation of the
  pseudo-momentum $P_{n_0 m_0}$, this Umklapp cannot relax the heat
  current alone, i.e. $\cal D$ in Eq.~(\ref{kappa-lowTh}) vanishes if
  only $n_0$ and $m_0$ are included in the sum. Hence $\kappa$ is 
determined by the second strongest Umklapp $H^U_{n_0'm_0'}$
with the smallest possible (but finite) momentum transfer, 
$\Delta k_{n_0' m_0'}=\pm G/n_0$ (the corresponding values of $n_0',m_0'$ depend
strongly on $n_0,m_0$). We therefore obtain for the heat conductivity close to commensurability,
\begin{eqnarray}\label{kappa-com}
\kappa_{com} \sim T^{4-2\alpha_{n_0'}}
\exp  \left[\frac{v_{2}G}{2n_0 T}\right]\; ,
\end{eqnarray}
where $\alpha_n$ is defined in Eq. (\ref{alpbet}).  The expression for
$\kappa_{com}$ is valid as long as $k_F$ is sufficiently close to the
commensurate value so that $\Delta k_{n_0m_0}\ll T/v_1$ (in which case
the ``high $T$'' approximation holds for $M_{n_0}^{pq}(\Delta
k_{n_0m_0},T)$). Note that the conductivity $\kappa$ is
 {\em largest} close to a
commensurate magnetization $M$ where $n_0$ is small in apparent
contradiction to the expectation that Umklapp is most efficient for
commensurate fillings. The reason is
simple\cite{rosch}: while the strongest Umklapp is enhanced for
commensurate $M$, the {\em second-strongest} which determines the size
of $\kappa$ is suppressed.
  
  How large is the conductivity $\kappa$ for a typical incommensurate
  magnetization or for temperatures where the asymptotic behavior
  (\ref{kappa-com}) is not yet reached? Umklapp processes $H^U_{nm}$
  with small $n$ are suppressed at low $T$ because $\Delta k_{nm}$ is
  large while contributions with large $n$ are suppressed by algebraic
  prefactors with large exponents $\alpha_n$ and $\beta_n$ [see
  Eqs.~(\ref{MhighT12}--\ref{MnlowT12})].
We therefore estimate $\kappa(T)$ using Eq.  (\ref{kappa-lowTh}) as
follows: first, the summation over $m$ is performed for a given $n$,
noting that the dominant term (which maximizes $M_{nm}$) is $m=m_0$,
where $m_0/n$ is the closest rational approximation of $2k_F/G$.
The {\em second-strongest} Umklapp (for this $n$) is therefore
  characterized by a momentum transfer $\Delta k\approx G/N$ where
  $N=n \alpha$ (with $\alpha$ of order unity). We then evaluate the
remaining sum over $n$ in a saddle point approximation.  This yields
\begin{eqnarray}\label{kappa-typ}
&&\kappa_{typ}(T) \sim \exp \!\left[\left(\tilde T^\ast/T\right)^{2/3}\right]
\end{eqnarray}
where $\tilde T^\ast = {\mathcal C}[\ln(v_{2}G/T)]^{1/2}v_{2}G$
and ${\mathcal C}$ is a constant of order unity. 

\begin{figure}
\begin{center}
\includegraphics[width=0.95 \linewidth,clip=]{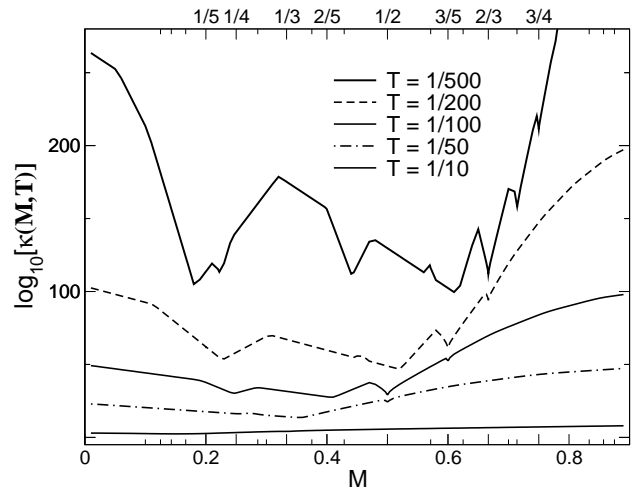}
\end{center}
\caption{\label{fig1}
Schematic plot of the dependence of the 
  logarithm of the heat conductivity $\kappa(M,T)$ on magnetization
  $M$ for various temperatures. $\kappa$ is largest for almost commensurate
  magnetizations and low temperatures. 
  Parameters used for this plot: $g^U_{nm}=1$, $G=2 \pi$, $v=1$, 
  $v_p=0.8$, $K=0.7$, $u=0.1 M$. }
\end{figure}

\begin{figure}
\begin{center}
\includegraphics[width=0.95 \linewidth,clip=]{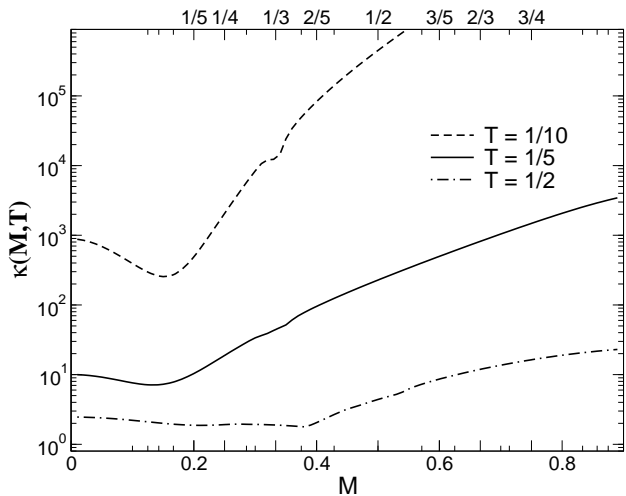}
\end{center}
\caption{\label{fig2}
Schematic plot of the magnetization dependence  of the heat conductivity 
$\kappa$  for various temperatures. Temperatures are higher than shown
in Fig.~\ref{fig1}, otherwise parameters are identical. } 
\end{figure}

 In Figs.~\ref{fig1} and \ref{fig2} we show schematically the
  dependence of $\ln[\kappa]$ on the magnetization. For a precise
  quantitative prediction of $\kappa(M,T)$ the RG flow of all Umklapp
  terms has to be calculated. While this is in principle possible e.g.
  for a highly anisotropic spin chain (small $J_z$) this is beyond the
  scope of this paper.  To obtain the schematic picture of
  Figs.~\ref{fig1} and \ref{fig2}, we have set all $g^U_{nm}$ to unity
  and used the asymptotic expressions (\ref{MhighT12}--\ref{MnlowT12})
  for the memory matrix in (\ref{kappa-lowTh}). As expected, the $M$
  dependence of $\kappa(M)$ becomes more and more ``spiky'' towards
  lower temperatures with maxima close to commensurate magnetizations,
  e.g. for $M=0,\frac{1}{3},\frac{1}{2},\frac{2}{3},..$.
  Unfortunately, the multiple peaks in $\kappa(M)$ occur at extremely
  high values of $\kappa$ where in most experimental systems the heat
  transport will be dominated by impurities or sample boundaries.
  Therefore this effect, while being an amusing theoretical
  prediction, is difficult to be observed experimentally.  At high
  temperatures, however, a pronounced minimum in $\kappa$ should be
  experimentally observable (see
  Fig.~\ref{fig2}) in a regime where inelastic scattering still dominates
  transport. The precise position of this minimum depends on the
  temperature and on details of the system under consideration.  We
  also would like to point out that a picture similar to
  Figs.~\ref{fig1} and \ref{fig2} would emerge even in the absence of
  spinon-phonon coupling, only the relevant energy scales would be
  different and set by $J$ rather than the minimum of $\Theta_D$ and
  $J$.  
  
Note that the position of the maxima in Fig.~\ref{fig1} seems to be
shifted away from the commensurate positions. Consider for example the
situation when $M$ is close to $\frac{1}{3}$, $M=\frac{1}{3}+\delta
M$. As $|\Delta k_{nm}|=G | n \frac{1+M}{2}-m|$ according to
Eqs.~(\ref{kf-h}) and (\ref{knm}), the dominating scattering process is
$H^U_{32}$ with $\Delta k_{32}=\frac{3 G}{2} \delta M$ being very small.
Therefore the heat transport is dominated by the relaxation of
$P_{32}$ by the second strongest Umklapp $H^U_{11}$ with $|\Delta
k_{11}|=G(\frac{1}{3}-\frac{1}{2}\delta M)$ and $\ln \kappa \approx
\frac{v_2 G}{2 T} (\frac{1}{3}-\frac{1}{2}\delta M)$ for small $\delta
M$ and low $T$. The position
of the local maximum to the left of $M=1/3$ is determined by the
competition with $H^U_{21}$ with $|\Delta k_{21}|=G
(\frac{1}{3}+\delta M)$. While the exponential factors in
Eqn.~(\ref{MnlowT12}) favor $H^U_{21}$ for $\delta M<0$ as $|\Delta
k_{21}|-|\Delta k_{11}|=\frac{3G}{2} \delta M<0$, the higher order
process $H^U_{21}$ is suppressed by algebraic prefactors and will
therefore only prevail for sufficient large $\delta M<0$ or sufficient
low $T$.

In comparison to Ref.~[\onlinecite{rosch}] it is interesting to note
that while the electrical (or spin-) conductivity in 1d systems is
suppressed for exactly commensurate fillings, this is not the case for
$\kappa$. The reason is that the overlap of the relevant
pseudo-momentum and the heat current is finite at commensurate
fillings, while the corresponding overlap of $P_{nm}$ and spin or
charge currents vanishes with exponential precision\cite{rosch02}
-- therefore also the Wiedemann-Franz law will be violated exponentially
for commensurate fillings.

\section{Conclusions}
\label{sec:conc}

In this paper we have studied the thermal conductivity $\kappa$ of
clean spin-chains coupled to phonons. The heat transport in the
absence of defects is strongly influenced by approximately conserved
pseudo-momenta. Due to their presence, low energy processes cannot
relax the heat current and therefore $\kappa$ is exponentially large
at low $T$. The exponent is determined by the slowest mode in the
system, i.e. in most materials by the phonon velocity. In
semi-quantitative agreement with experiments\cite{spin-ex-umk} we find
$\kappa \sim \exp[ \Theta_D/(2 c T)]$ where $\Theta_D$ is the Debye
temperature and $c=\frac{2 \Theta_D}{v_p G}$ a number close to 1
depending on the anisotropies of the phonon system. 
It is instructive to compare this to the pure phonon Umklapp which
leads to a relaxation of the phonon contribution to the heat current
characterized by twice as large an exponent, $\kappa \sim \exp[-v_p
G/(2 T)]$. The reason is that in the mixed spinon-phonon process
considered by us the relevant momentum transfer is $G-2 k_F=G/2$ rather 
than $G$. Indeed, by comparing the plots of $\ln \kappa(T)$ 
vs. $1/T$ of the phonon and the spinon contribution in
Sr$_2$CuO$_3$\cite{solo} one can identify slopes differing by this
factor of $2$.

In the presence of a magnetic field, spinon and phonon modes start to
mix. Furthermore, the Umklapp processes depend exponentially on the
magnetization. This leads to a fractal-like spiky dependence of
$\kappa$ on magnetization. Surprisingly, $\kappa$ is largest for
commensurate magnetizations. This is again a consequence of the
approximately conserved pseudo-momenta: it is not the strongest, but
the second strongest Umklapp process which determines the thermal
transport.

\acknowledgments We would like to thank B. B\"uchner, 
S.W. Cheong, L.B.~Ioffe, M.R. Li, E. Orignac,
 A.V. Sologubenko, D.~Vanderbilt, Y.J. Wang and  P. W\"olfle
for helpful and enlightening
 discussions.

\begin{appendix}

\section{Conductivity, approximate conservation laws and boundary conditions} 
\label{appendixAppr}

In this appendix we discuss on general grounds, how boundary
conditions and approximate conservation laws influence transport
measurements. Let us assume that in an experiment the transport of a
conserved charge $q_1$ (e.g. the energy density) is measured using a
4-point probe. We want do investigate the influence of a second
approximately conserved charge $q_2$ (e.g. the spin-density) on the
experiment. The arguments given below can easily be generalized to
include further exact or approximate conserved charges. The relevant
continuity equations read
\begin{eqnarray}
\partial_t q_1+\partial_x j_1 &=&0\label{Ac1}\\
\partial_t q_2+\partial_x j_2 &=&-q_2/\tau_2,\label{Ac2}
\end{eqnarray}
where $\tau_2$ describes phenomenologically the (slow) relaxation of
$q_2$ (assuming that $q_2=0$ in equilibrium by definition). To set up
a hydrodynamic description of the measurement, we assume that currents
are driven by the force $F_1$ (e.g. $\partial_x T/T$) and by
gradients of $q_2$
\begin{eqnarray}
j_1&=&\sigma_{11} F_1 - D_{12} \partial_x q_2 \label{Aj1}\\
j_2&=&\sigma_{21} F_1 - D_{22} \partial_x q_2,\label{Aj2}
\end{eqnarray} 
where the conductivity $\sigma_{12}$ describes e.g. the spin-analog of
 thermopower. In steady state, $\partial_t q_i=0$, one obtains
easily from Eqs.~(\ref{Ac1}--\ref{Aj2})
\begin{eqnarray}
j_2&=&j_1 \sigma_{21}/\sigma_{11}-\tilde{D}_{22} \partial_x q_2\label{a5}\\
\tilde{D}_{22} \partial_x^2 q_2&=&q_2/\tau_2 \label{a6}
\end{eqnarray}
with $j_1(x)=const.$ and $\tilde{D}_{22}=D_{22}-D_{12}
\sigma_{21}/\sigma_{11}$.  These equations have to be solved with the
appropriate boundary conditions. For our example the experimentally
relevant boundary conditions are $j_2(\pm L/2)=0$ (no spin-current
flowing out of the sample) where $L$ is the size of the sample. From (\ref{a6})
one therefore obtains for $-L/2\le x \le L/2$
\begin{eqnarray}
q_2=q_2^0 \left(\exp\!\left[-\frac{x+L/2}{l_2}\right]-
\exp\!\left[\frac{x-L/2}{l_2}\right]\right)
\end{eqnarray}
where $q_2^0$ is determined from the boundary conditions and 
$l_2=\sqrt{\tilde{D}_{22} \tau}$ is the (spin-) diffusion length.

Obviously, one has to distinguish two different situations. If
$L\gg l_2$ both $q_2$ and $\partial_x q_2$ vanishes exponentially in the
sample. Therefore according to (\ref{Aj1}) the (heat) transport of
$q_1$ inside the sample is determined by $\sigma_{11}$.
\begin{equation}
j_1 = \sigma_{11} F_1 \quad \text{for } L\gg l_2.
\end{equation}
This is the situation considered in this paper (formally, within our
Hamiltonian $l_2=\infty$ as $S_z$ is conserved, but in a real
material $S_z$ will decay e.g. due to spin-orbit scattering from
impurities). Note that Eq.~(\ref{Aj2}) implies that the (spin-)
current $j_2$ inside the sample is finite for $\sigma_{12}\neq 0$,
i.e. for $h\neq0$. This current, however, decays close to the sample
boundaries in a width of order $l_2$.

The situation is different in the second case when $L\ll l_2$ or if
$q_2$ is exactly conserved. Then $j_2$ vanishes inside the sample due
to the boundary conditions and plugging (\ref{Aj2}) into (\ref{Aj1})
one obtains
\begin{equation}
j_1 = \left(\sigma_{11}-\frac{D_{12} \sigma_{21}}{D_{22}}\right) F_1  \quad \text{for } L\ll l_2.
\end{equation}

In passing, we note that in general the presence of approximate
conservation laws implies the existence of large length scales like $l_2$
on which transport is inhomogeneous. This may be related to the
experimental observation\cite{buechner}
 that heat transport in spin chains is often
extremely inhomogeneous.

\section{The calculations} \label{appendixCalc}
To evaluate the matrix elements $M_{nml}^{pq}$, we first find explicit
expressions for the operators $F^p_{nml}(t)=i[J_p,H^U_{nml}]$
($l=0,1$), where $H^U_{nm0}=H^U_{nm}$, $H^U_{nm1}=H^{U,s-p}_{nm}$ and
the currents are given by Eqs. (\ref{Js}), (\ref{JQ}). We restrict our
calculation to the even $n$ terms (\ref{Unm-sc}), (\ref{Unm-sp}). The
extension to odd $n$ [Eqs. (\ref{Unm-scOdd}) and (\ref{Unm-spOdd})] is
straightforward: due to the extra factors of $\partial_x \phi$, it
will essentially amount to trading $C_{ss}$ by $C_{dd}$ below [see
Eqs. (\ref{def-Css0}), (\ref{def-CQQ0})].  We also define all
operators below for a single chain.  To leading order in $(v_p/v)\ll
1$, we find
\begin{eqnarray}\label{Fnmlp}
F_{nml}^s(t)&=& \frac{-ig^U_{nml}2nKv}{(2 \pi a)^n}\int dx
(e^{i \Delta k x} e^{ i  2 n \phi(x,t) } b_l(x) - h.c.)\; , \nonumber \\
F_{nml}^Q(t)&\approx & \frac{ig^U_{nml}2nv^2}{(2 \pi a)^n} \\ \,& \times & \int dx 
(e^{i \Delta k x} e^{ i  2 n \phi(x,t) }\partial_x \phi b_l(x) - h.c.)\; , \nonumber 
\end{eqnarray}
where for abbreviation we have omitted the indices $n,m$ from $\Delta k_{nm}$ and 
introduced the definition 
\begin{eqnarray}\label{bx}
b_l(x)\equiv \left(\partial_x q \right)^l\; .
\end{eqnarray}
We then find that the retarded correlation function 
$\langle F_{nml}^p;F^q_{nml}\rangle^0_{\w}$ at frequency $\w$ is given by \cite{schultz}
\begin{eqnarray}\label{retcor}
\langle F_{nml}^p;F^q_{nml}\rangle^0_{\w}=&\, & \\
2A_{pq}\int_{-\infty}^{\infty}dx\int_{0}^{\infty} dt
& e^{i(\w t-\Delta k x)}& {\rm Im}\{C_{pq}(x,t)\} \; ,\nonumber 
\end{eqnarray}
where 
\begin{eqnarray}\label{def-Css0}
C_{ss}(\xi)&=&\langle \exp[i2n\phi(\xi)]\exp[-i2n\phi(0)]\rangle^0{\mathcal G}_l(\xi)
\end{eqnarray}
and
\begin{eqnarray}\label{def-CsQ0}
C_{sQ}(\xi)&=&C_{Qs}(\xi)=v^2  C_d(\xi){\mathcal G}_l(\xi) \\
C_{QQ}(\xi)&=&v^4  C_{dd}(\xi) {\mathcal G}_l(\xi)\; , \label{def-CQQ0} \\
C_d(\xi)&\equiv &\langle \partial_x \phi(\xi)
\exp[i2n\phi(\xi)]\exp[-i2n\phi(0)]\rangle^0\nonumber \\
C_{dd}(\xi)&\equiv &
\langle \partial_x \phi(\xi)\partial_x \phi(0) 
\exp[i2n\phi(\xi)]\exp[-i2n\phi(0)]\rangle^0 \nonumber
\end{eqnarray}
($\xi$ is an abbreviation for $(x,t)$);
\begin{eqnarray}\label{Glprop}
{\mathcal G}_l(\xi)\equiv \langle b_l^\dagger(\xi)b_l(0) \rangle
\end{eqnarray}
\begin{eqnarray}\label{Aqp}
A_{ss}&=&\frac{4(g^U_{nml})^2(nKv)^2}{(2 \pi a)^{2n}}\; , \nonumber \\
A_{sQ}&=&A_{Qs}=\frac{4(g^U_{nml})^2n^2Kv}{(2 \pi a)^{2n}}\; ,\\
A_{QQ}&=&\frac{4(g^U_{nml})^2n^2}{(2 \pi a)^{2n}}\; .\nonumber 
\end{eqnarray}
Eq. (\ref{def-Css0}) yields
\begin{eqnarray}\label{Css}
C_{ss}(x,t)&=&e^{4n^2G_\phi(x,t)}{\mathcal G}_l(x,t) \; , \\
G_\phi(x,t)&\equiv & \langle  \phi(x,t)\phi(0,0)\rangle^0 \nonumber 
\end{eqnarray}
where at finite $T$ the Green's function $G_\phi(x,t)$ is given by \cite{schultz}
\begin{eqnarray}\label{Green-T0}
 G_\phi(x,t)&=& \frac{K}{4} \ln \left[\frac{\pi aT/v}
{\sinh\{\pi T(x-v t+ia)/v\}}\right] \\
\, &+& \frac{K}{4}\ln \left[\frac{\pi aT/v}
{\sinh\{\pi T(x+v t-ia)/v\}}\right]\; . \nonumber 
\end{eqnarray}
${\mathcal G}_0(\xi)=1$, and the phonon propagator ${\mathcal G}_1(\xi)=G_p(\xi)$
(at finite $T$) has the form 
\begin{eqnarray}\label{freeGl}
G_p(x,t)&\approx & B\left[\frac{\pi aT/v_p}
{\sinh\{\pi T(x-v_p t+ia)/v_p\}}\right] \nonumber \\
& \times & \left[\frac{\pi aT/v_p}
{\sinh\{\pi T(x+v_p t-ia)/v_p\}}\right]
\end{eqnarray}
with $B$ a numerical constant. 
\begin{widetext}
Substituting in Eq. (\ref{Css}) and using the notation
$B_l$, where $B_0=1$ and $B_1=B$, we obtain
\begin{multline}\label{corr-xt0}
C_{ss}(x,t)= B_l\left(\frac{\pi aT}{v}\right)^{2n^2K}
\left(\frac{\pi aT}{v_p}\right)^{2l} 
[\sinh\{\pi T(x - v t+ia)/v\}\sinh\{\pi T(x+v t-ia)/v\}]^{-n^2K}\\
\times
[\sinh\{\pi T(x - v_p t+ia)/v_p\}
\sinh\{\pi T(x+v_p t-ia)/v_p\}]^{-l}\, . 
\end{multline}
Using the identity 
\begin{equation}
\partial_x \phi(x,t)=\lim_{\gamma\rightarrow 0}
\lim_{y\rightarrow x}(i\gamma)^{-1}\partial_y\exp[i\gamma\phi(y,t)] \nonumber
\end{equation}
we can also express $C_d$, $C_{dd}$ in terms of the function
$G_\phi(x,t)$ and its derivatives:
\begin{eqnarray}\label{Cd-xt}
C_d(x,t)&=& -2n\partial_xG_\phi(x,t)C_{ss}(x,t) \\
C_{dd}(x,t)&=& 
 C_{ss}(x,t)\left[4n^2(\partial_xG_\phi(x,t))^2+
\partial_x^2G_\phi(x,t)\right] , \label{Cdd-xt}
\end{eqnarray}
where
\begin{eqnarray}\label{Green-1der}
\partial_x G_\phi(x,t)&=&-\frac{K}{4} \frac{\pi T}{v}
[\coth\{\pi T(x-v t+ia)/v\}+ 
\coth\{\pi T(x+v t-ia)/v\}]
\end{eqnarray}
and
\begin{eqnarray}\label{Green-2der}
\partial_x^2 G_\phi(x,t)&=&\frac{K}{4} 
\left(\frac{\pi T}{v}\right)^2
[\sinh^{-2}\{\pi T(x-v t+ia)/v\}+ 
\sinh^{-2}\{\pi T(x+v t-ia)/v\}].
\end{eqnarray}

We now recall Eq. (\ref{MMM}) for $M_{nml}^{pq}$, and consider the limit 
$\w\rightarrow 0$ where the correlation functions
$\langle F_{nml}^p;F^q_{nml}\rangle^0_{\w}$ given by Eq. (\ref{retcor}) are expanded
to linear order in $\w$. We then get
\begin{eqnarray}\label{MMM-w0}
M_{nl}^{pq}(\Delta k,T)&=&\lim_{\w\rightarrow 0}M_{nml}^{pq} 
=A_{pq}\int_{-\infty}^{\infty}dx e^{-i\Delta k x}\int_{-\infty}^{\infty} dt\, t
 {\rm Im}\{C_{pq}(x,t)\},
\end{eqnarray}
where we have used the fact that $C_{pq}^\ast(x,t)=C_{pq}(x,-t)$ [see Eqs.
(\ref{retcor}) through (\ref{Green-2der})] and hence the function 
$t {\rm Im}\{C_{pq}(x,t)\}$ is symmetric with respect to $t\rightarrow -t$.
The matrix element $M_{nl}^{ss}(\Delta k,T)$ can be computed by substituting
Eq. (\ref{corr-xt0}) in (\ref{MMM-w0}) and performing the integrals. To compute
$M_{nl}^{sQ}(\Delta k,T)$, $M_{nl}^{QQ}(\Delta k,T)$ we insert  Eqs. (\ref{Aqp}),
(\ref{corr-xt0}), (\ref{Green-1der}) and (\ref{Green-2der}) into (\ref{Cd-xt}), 
(\ref{Cdd-xt}). This yields  
\begin{eqnarray}\label{CsQ-xt0}
A_{sQ}C_{sQ}(x,t)&\approx&
\frac{2(g^U_{nml})^2}{(2 \pi a)^{2n}}\frac{(nv)^3K^2B_l}{a} 
\left(\frac{\pi aT}{v}\right)^{2n^2K+1}
\left(\frac{\pi aT}{v_p}\right)^{2l} f_{sQ}(x,t)
\end{eqnarray}
\begin{multline}
f_{sQ}(x,t)\equiv  [\coth\{\pi T(x-v t+ia)/v\}+
\coth\{\pi T(x+v t-ia)/v\}] \left[\sinh\{\pi T(x-v t+ia)/v\}\sinh\{\pi T(x+v t-ia)/v\}\right]^{-n^2K}\\
\times \left[\sinh\{\pi T(x - v_p t+ia)/v_p\}
\sinh\{\pi T(x+v_p t-ia)/v_p\}\right]^{-l} ,
\nonumber 
\end{multline}
and
\begin{eqnarray}\label{CQQ-xt0}
A_{QQ}C_{QQ}(x,t)&\approx&
\frac{(g^U_{nml})^2}{(2 \pi a)^{2n}}\frac{v^4n^2KB_l}{a^2} \left(\frac{\pi aT}{v}\right)^{2n^2K+2}
\left(\frac{\pi aT}{v_p}\right)^{2l} f_{QQ}(x,t) , 
\end{eqnarray}
\begin{multline}
f_{QQ}(x,t)\equiv  \bigl[
n^2K(\coth\{\pi T(x-v t+ia)/v\}+ \coth\{\pi T(x+v t-ia)/v\})^2 + 
\sinh^{-2}\{\pi T(x-v t+ia)/v\}  \\ + 
\sinh^{-2}\{\pi T(x+v t-ia)/v\} \bigr] 
\left[\sinh\{\pi T(x-v t+ia)/v\}\sinh\{\pi T(x+v t-ia)/v\}\right]^{-n^2K}
\nonumber \\ 
\times \left[\sinh\{\pi T(x-v_p t+ia)/v_p\}\sinh\{\pi T(x+v_p t-ia)/v_p\}\right]^{-l} .
\nonumber 
\end{multline}
\end{widetext}

The last step is to insert the functions given by Eqs. (\ref{corr-xt0}),
(\ref{CsQ-xt0}) and (\ref{CQQ-xt0}) into Eq. (\ref{MMM-w0}) and evaluate the integrals.
We employ the change of variables $s=T(x/v_p-t)$, $s^\prime=T(x/v_p+t)$, and define
a dimensionless parameter $\lambda\equiv v_p|\Delta k|/T$. Eq. (\ref{MMM-w0}) is
then recast in the form
\begin{eqnarray}\label{Mnlambda}
&&M_{nl}^{pq}(\Delta k,T)=T^{\eta_{pq}+2(n^2K+l)-3} \\ &&\times 
\int_{-\infty}^{\infty}\int_{-\infty}^{\infty} ds ds^\prime
e^{-i\, \text{sgn}(\Delta k)\lambda(s+s^\prime)/2}{\mathcal F}_{nl}^{pq}(s,s^\prime)
\nonumber
\end{eqnarray}
where $\eta_{ss}=0$, $\eta_{sQ}=1$, $\eta_{QQ}=2$ and the functions
${\mathcal F}_{nl}^{pq}(s,s^\prime)$ do not contain any dependence on
$\Delta k$ and $T$.  Hence the double integral depends on them only
through the parameters $\lambda$ and $\text{sgn}(\Delta k)$ in the
exponential factor.  In particular, the high and low $T$ limit cases
are distinguished by $\lambda\ll 1$ and $\lambda\gg 1$, respectively.
In the high $T$ limit the exponential factor is expanded up to first
order in $\lambda$.  The leading contribution to $M_{nl}^{ss}$ and
$M_{nl}^{QQ}$ comes from the $0$'th order, i.e. the integration
results in a constant independent of $\Delta k$ and $T$. However,
since $C_{sQ}(x,t)$ is an odd function of $x$, the leading
contribution to $M_{nl}^{sQ}$ comes from the first order leading to an
overall factor of $\Delta k/T$.  These approximations yield Eq.
(\ref{MhighT}).

To get the low $T$ limit expressions for  $M_{nl}^{pq}(\Delta k,T)$, we evaluate the
integrals in Eq. (\ref{Mnlambda}) in a saddle point approximation where the 
large parameter is $\lambda$. As implied by Eqs. (\ref{corr-xt0}), (\ref{CsQ-xt0}) and 
(\ref{CQQ-xt0}), the functions $C_{pq}(x,t)$ and hence 
${\mathcal F}_{nl}^{pq}(s,s^\prime)$ have branch--cut
singularities along the imaginary axis (this is for the generic case where
$K$ is not an integer), one of which is close to the real axis.
The integration over $s^\prime$ is performed first, with a slight 
deformation of the real axis to include the single saddle point 
\begin{eqnarray}
s^\prime_0=-i\frac{2}{\lambda}\approx i0^{-}\; .
\end{eqnarray} 
Then, for $\text{sgn}(\Delta k)<0$ ($\text{sgn}(\Delta k)>0$), the contour of
integration over $s$ is deformed from the real axis to the upper (lower) imaginary axis. 
The series of saddle points $\{s_i\}_{i=0}^{\infty}$
dominating this integral are close to the zeros of the
$\sinh$ functions in (\ref{corr-xt0}) up to a correction of order 
$1/\lambda\rightarrow 0$. Since the contribution of each saddle point 
involves an exponential factor $e^{-\lambda |s_i|/2}$, the overall result
will be dominated by the minimal $s_i$ for which the functions
${\mathcal F}_n^{pq}(s,s^\prime)$ do not vanish by symmetry. The latter requirement
excludes the contribution of $s_0=s^\prime_0$. The leading contribution therefore 
originates from the saddle point
\begin{eqnarray}
s_{1}=\pm i\, ,
\end{eqnarray}
where $\text{sgn}(\Delta k)=\mp$. Noting once again that $v_p\ll v$, we  obtain 
the low $T$ expression for $M_{nl}(\Delta k,T)$ in Eq. (\ref{MnlowT}). The matrix element
$M_{nl}^{ss}(\Delta k,T)$ is then given directly up to a prefactor. To
get the other matrix elements in (\ref{MlowT}), we use scaling arguments noting that
as long as $(T/v |\Delta k|) \ll (v_p/v)$, the dominant momentum scale is $\Delta k$.

The derivation of $M_{n}^{pq}(\Delta k,T)$ [Eqs. (\ref{MhighT12}) through 
(\ref{MnlowT12})] in the case where spinons and phonons couple linearly involves 
essentially the same calculation. In this case, however, the memory matrix is dominated by 
the contribution of the $l=0$ Umklapp terms Eq. (\ref{Unm-sc}), (\ref{Unm-scOdd}). 
In addition, the spinon--phonon mixing introduces
the fields $\phi_1$, $\phi_2$ (the eigenmodes of the Hamiltonian $H^*$), in terms of which
the spinon field $\phi$ can be written as
\begin{equation}\label{phito12}
\phi=\sqrt{K}\left(C\phi_1-S\left(\frac{v}{v_p}\right)^{1/2}\phi_2\right) 
\end{equation}
where we have used the definition $\phi=\sqrt{K}\tilde\phi$ and the
transformation Eq. (\ref{phi1phi2}). As a consequence the operators
$F_{nm}^p=i[J_p,H^U_{nm}]$ (for even $n$) are given by
\begin{eqnarray}\label{Fnmp}
F_{nm}^s(t)&=& \frac{-ig^U_{nm}2nv}{(2 \pi a)^n}\int dx
(e^{i \Delta k x} e^{ i  2 n \phi(x,t) } - h.c.)\; , \nonumber \\
F_{nm}^Q(t)&=& \frac{ig^U_{nm}\sqrt{K}2n}{(2 \pi a)^n}\int dx
(e^{i \Delta k x} e^{ i  2 n \phi(x,t) } - h.c.) \nonumber \\ \, & \, & \times
\left(v_1^2 C\partial_x \phi_1-
v_2^2 S\left(\frac{v}{v_p}\right)^{1/2}\partial_x \phi_2\right)\, .  
\end{eqnarray}
The correlators $\langle F_{nm}^p;F^q_{nm}\rangle^0_{\w}$ are again written in terms of the
functions $C_{pq}(x,t)$  [Eq. (\ref{retcor})]. Here, 
\begin{eqnarray}\label{def-Css}
C_{ss}(\xi)&=&\langle \exp[i2n\phi(\xi)]\exp[-i2n\phi(0)]\rangle^0
\end{eqnarray}
and
\begin{eqnarray}\label{def-CsQ}
C_{sQ}(\xi)&=&C_{Qs}(\xi)=v_1^2 C C_1(\xi)-
v_2^2 S\left(\frac{v}{v_p}\right)^{1/2}C_2(\xi)\nonumber \\
C_{QQ}(\xi)&=&v_1^4 C^2 C_{11}(\xi)-
2(v_1v_2)^2 CS\left(\frac{v}{v_p}\right)^{1/2}C_{12}(\xi) \nonumber \\
\,&+&(v_2)^4 S^2\frac{v}{v_p}C_{22}(\xi), \label{def-CQQ} \\   
C_\nu(\xi)&\equiv &\langle \partial_x \phi_\nu(\xi)
\exp[i2n\phi(\xi)]\exp[-i2n\phi(0)]\rangle^0\nonumber \\
C_{\nu\nu^\prime}(\xi)&\equiv &
\langle \partial_x \phi_\nu(\xi)\partial_x \phi_{\nu^\prime}(0) 
\exp[i2n\phi(\xi)]\exp[-i2n\phi(0)]\rangle^0 \nonumber
\end{eqnarray}
and the coefficients $A_{pq}$ are given by
\begin{eqnarray}\label{Aqph}
A_{ss}&=&\frac{4(g^U_{nm})^2(nKv)^2}{(2 \pi a)^{2n}}\; , \nonumber \\
A_{sQ}&=&A_{Qs}=\frac{4(g^U_{nm})^2n^2vK^{3/2}}{(2 \pi a)^{2n}}\; ,\\
A_{QQ}&=&\frac{4(g^U_{nm})^2n^2K}{(2 \pi a)^{2n}}\; .\nonumber 
\end{eqnarray}
Inserting Eq. (\ref{phito12}) into (\ref{def-Css}) we get
\begin{eqnarray}\label{Css12}
C_{ss}(x,t)&=&e^{4n^2KC^2G_1(x,t)} e^{4n^2KS^2(v/v_p)G_2(x,t)},  \nonumber \\
G_\nu(x,t)&\equiv & \langle  \phi_\nu(x,t)\phi_\nu(0,0)\rangle^0 
\end{eqnarray}
where similarly to Eq. (\ref{Green-T0})
\begin{eqnarray}\label{Green-T}
 G_\nu(x,t)&=& \frac{K_\nu}{4} \ln \left[\frac{\pi aT/v_\nu}
{\sinh\{\pi T(x-v_\nu t+ia)/v_\nu\}}\right] \\
\, &+& \frac{K_\nu}{4}\ln \left[\frac{\pi aT/v_\nu}
{\sinh\{\pi T(x+v_\nu t-ia)/v_\nu\}}\right]\; . \nonumber 
\end{eqnarray} 
Also, similarly to the derivation of Eqs. (\ref{Cd-xt}), (\ref{Cdd-xt}) we obtain 
expressions for $C_{\nu}$, $C_{\nu\nu^\prime}$ in terms of the functions 
$G_\nu(x,t)$ and their derivatives:
\begin{eqnarray}\label{C1C2}
C_1(x,t)&=& -2n\sqrt{K}C\partial_xG_1(x,t)
C_{ss}(x,t)  \\
C_2(x,t)&=& -2n\sqrt{K}S\left(\frac{v}{v_p}\right)^{1/2}
\partial_xG_2(x,t)C_{ss}(x,t), \nonumber
\end{eqnarray}
\begin{eqnarray}\label{C11C22}
C_{11}(x,t)&=& C_{ss}(x,t)\left[4n^2KC^2(\partial_xG_1(x,t))^2+\partial_x^2G_1(x,t)\right] \nonumber \\
C_{12}(x,t)&=& 4n^2KCS\left(\frac{v}{v_p}\right)^{1/2}\times 
\nonumber \\ \, &\times & 
\partial_xG_1(x,t)\partial_xG_2(x,t)C_{ss}(x,t) \\
C_{22}(x,t)&=& C_{ss}(x,t) \nonumber \\ \, &\times &
\left[4n^2KS^2\frac{v}{v_p}
\left(\partial_xG_2(x,t)\right)^2+\partial_x^2G_2(x,t)\right]. \nonumber
\end{eqnarray}
The explicit dependence on $x$ and $t$ is obtained from Eqs.
(\ref{Green-T}), (\ref{Css12}), which yield
\begin{multline}\label{corr-xt}
C_{ss}(x,t)= \left(\frac{\pi aT}{v_1}\right)^{2\alpha_n}
\left(\frac{\pi aT}{v_2}\right)^{2\beta_n} \times \\
[\sinh\{\pi T(x - v_1 t+ia)/v_1\}\sinh\{\pi T(x+v_1 t-ia)/v_1\}]^{-\alpha_n}
 \\ [\sinh\{\pi T(x - v_2 t+ia)/v_2\}
\sinh\{\pi T(x+v_2 t-ia)/v_2\}]^{-\beta_n} 
\end{multline}
$\alpha_n=n^2KK_1C^2$ and $\beta_n=n^2KK_2S^2(v/v_p)$.
Note that this expression is essentially the same function as (\ref{corr-xt0}) with different
parameters: $v\rightarrow v_1$, $v_p\rightarrow v_2$, $n^2K\rightarrow\alpha_n$ and
$l\rightarrow \beta_n$. 
As a result, the integrals in Eq. (\ref{retcor}) can be evaluated in the
same manner, yielding Eqs. (\ref{MhighT12}) through (\ref{MnlowT12}). 

\section{Erratum}
\label{erratum}

\begin{widetext}

During the preparation of a subsequent article \cite{BMARS}, we have spotted 
an error in the derivation of the effective propagator of 3-dimensional phonons
$G_p(t,x)$, obtained after integration over the momenta in the direction 
perpendicular to the spin chains. The revised calculation of the thermal 
conductivity $\kappa(h=0)$ as a function of $T$ [Eq. (47)] yields a modified 
power-law prefactor, however, our main result -- the exponential behavior with
the characteristic temperature scale $T^*$ [Eq. (48)] -- is unchanged.

The effective single--phonon propagator 
$G_p(x,t)=\langle\partial_x q(x,t)\partial_x q(0,0)\rangle $ (where $q$ is
the displacement field in a particular chain) can be derived 
by an inverse Fourier transformation of 
$G_p(k,t)=k^2\int d^2k_{\perp} G_p^{3D}(\vec{k},t)$, where $k$ is the 
component of $\vec{k}$ along the chain and $G_p^{3D}(\vec{k},t)$ the 
correlator $\langle q_{\vec{k}}(t)q_{\vec{k}}(0)\rangle $ 
of free (isotropic) $3D$--phonons at finite $T$:
\begin{equation}
G_p^{3D}(\vec{k},t>0)=const \times \frac{1}{|\vec{k}|}
\left[n_{\vec{k}}e^{iv_p|\vec{k}|t}+
\left(1+n_{\vec{k}}\right)e^{-iv_p|\vec{k}|t}\right]\; ,
\end{equation}
where $n_{\vec{k}}=(e^{v_p|\vec{k}|/T}-1)^{-1}$ is the phonon occupation.
We obtain the following expression, to replace Eq. (B11) in Appendix B:
\begin{multline}\label{freeGlcorr}
G_p(x,t)=  B\bigl[\frac{1}{v_pt}\left(\frac{1}{(x+v_pt)^3}-
\frac{1}{(x-v_pt)^3}\right) \\ +\left(\frac{T}{v_p}\right)^4
\sum_{n=1}^\infty\sum_{\sigma =\pm}
\left(\frac{1}{(n+\sigma itT)(n+\sigma i(t-x/v_p)T)^3}
+\frac{1}{(n+\sigma itT)(n+\sigma i(t+x/v_p)T)^3}\right) \bigr]\; .
\end{multline}
with $B$ a numerical constant. 

To find the contribution $M_{nl}^{pq}(\Delta k,T)$ to the memory
matrix (where $l\not =0$ denotes the number of phonons involved in the
Umklapp process), we use ${\mathcal G}_l(\xi)=[G_p(\xi)]^l$ in Eq.
(B9) with $G_p$ given by Eq. (\ref{freeGlcorr}) above. At low $T$, the
single--phonon process $l=1$ yields a {\em subdominant} contribution
$\sim e^{-v_p|\Delta k|/T}$. The physical reason for that is that the
energy cost of the process involves an excitation of a single phonon
that carries the entire momentum transfer $\Delta k$, i.e.
$v_p|\Delta k|$. It is energetically favorable to distribute the
momentum equally between the initial and final state using a
two--phonon process ($l=2$): a phonon with momentum $\Delta k/2$ is
scattered to a final state with momentum $-\Delta k/2$.  This yields
the leading contribution to the matrix element $M^{TT}$:
\begin{equation}
M_{12}^{TT}\sim T^{2K+3}\exp \left[-\frac{v_p |\Delta k|}{ 2T} \right]\; ,
\end{equation}
which should replace $M_{11}^{TT}$ in Eq. (46) (with $\Delta k=G/2$).
Higher order processes ($l>2$) give the same exponential
dependence but are supressed by power law prefactors due to phase
space restrictions.  The resulting expression for the thermal
conductivity is
\begin{eqnarray}
\kappa(h=0) &\approx & \kappa_0 \left(\frac{T^\ast}{T}\right)^{2K}
\exp  \left[\frac{T^\ast}{ T}\right]\; ,\quad T^\ast=\frac{v_pG}{4}\; .
\end{eqnarray}
This should replace Eq. (47).
\end{widetext}

\end{appendix}


\end{document}